# Crystallographic reconstruction study of the effects of finish rolling temperature on the variant selection during bainite transformation in C-Mn high-strength steels

Nicolas Bernier[1], Lieven Bracke[1], Loïc Malet[2], Stéphane Godet[2]

[1] OCAS N.V., ArcelorMittal R&D Gent, Pres. J.F. Kennedylaan 3, 9060 Zelzate, Belgium

[2] Université Libre de Bruxelles, 4 MAT (Materials Engineering, Characterisation, Synthesis and Recycling), Avenue F.D. Roosevelt 50, 1050 Brussels, Belgium

n.bernier@yahoo.fr (N. Bernier)

lieven.bracke@arcelormittal.com; tel: +32(0)93451374 (L. Bracke)

**Abstract**

The effect of finish rolling temperature (FRT) on the austenite-($\gamma$) to-bainite ($\alpha$) phase transformation is quantitatively investigated in high-strength C-Mn steels. In particular, the present study aims to clarify the respective contributions of the $\gamma$ conditioning during the hot rolling and the variant selection (VS) during the phase transformation to the inherited texture. To this end, an alternative crystallographic $\gamma$ reconstruction procedure, which can be directly applied to experimental electron backscatter diffraction (EBSD) mappings, is developed by combining the best features of the existing models: the orientation relationship (OR) refinement, the local pixel-by-pixel analysis and the nuclei identification and spreading strategy. The applicability of this method is demonstrated on both quenching and partitioning (Q&P) and as-quenched lath-martensite steels. The results obtained on the C-Mn steels confirm that the sample finish rolled at the lowest temperature (829°C) exhibits the sharpest transformation texture. It is shown that this sharp texture is exclusively due to a strong VS from parent brass {110}<1-12>, S {213}<-3-64> and Goss {110}<001> grains, whereas the VS from the copper {112}<-1-11> grains is insensitive to the FRT. In addition, a statistical VS analysis proves that the habit planes of the selected variants do not systematically correspond to the predicted active $\gamma$ slip planes using the Taylor model. In contrast, a correlation between the Bain group to which the selected variants belong and the FRT is clearly revealed, regardless of the parent orientation. These results are discussed in terms of polygranular accommodation mechanisms, especially in view of the observed development in the hot-rolled samples of high-angle grain boundaries with misorientation axes between $<111>_\gamma$ and $<110>_\gamma$.

# 1. Introduction

High-strength bainitic steels in hot worked conditions have been developed for a number of applications in a variety of industries, such as the automotive [1], railway [2], heavy machinery and general construction industries, because they combine superior tensile strength levels with improved toughness, ductility, formability and weldability using a low amount of carbon or other alloying elements [3]. Thermomechanical controlled processing (TMCP) is a common way to obtain this combination of mechanical properties [4]. It consists of slab reheating at well-defined temperatures followed by controlled rolling in the unrecrystallised $\gamma$ temperature range, i.e., below the non-recrystallisation temperature ($T_{nr}$). Under these circumstances, sufficient strain is accumulated in $\gamma$, resulting in a fairly sharp rolling texture and the formation of structural defects (in-grain shear bands, deformation bands, mechanical twins, etc.) acting as preferential nucleation sites for the transformation products. Combined with an accelerated cooling, this process can produce low-C steels exhibiting substantial grain refinement [5-6], which is one of the most effective ways to improve both strength and toughness. The presence of carbides and secondary phases, the density of dislocations and the transformation texture also affect the final mechanical properties of bainitic steels, e.g., the fracture behaviour [7-9]. The inheritance of these microstructure details depends on both the $\gamma$ conditioning during hot forming and the $\gamma$-to-$\alpha$ phase transformation. It is therefore crucial to investigate the effects of various TMCP parameters on the $\gamma$ conditioning and phase transformation to fine-tune the current processing route for plates and coils, improving upon the current strength and toughness limits.

The phase transformation during cooling prevents the direct observation of the $\gamma$ microstructure and texture. To overcome this issue, several "indirect" methods have been described in the literature. Some studies rely on traditional metallographic techniques [10-11], but they only provide morphological data, such as the prior $\gamma$ grain size. Furthermore, they have proven largely ineffective for steel compositions with a low concentration of residual elements (e.g., P) and/or reconstructive (i.e., diffusive) phase transformation products [12-13]. Other methods consist of comparing the texture measured on the transformed phases with those predicted by assuming a parent texture and a specific orientation relationship (OR) [14-17]. These assumptions are necessary, as more than one $\gamma$ orientation can transform into the same $\alpha$ orientation [18]. The lack of a known starting texture is clearly a major limitation to the reliability of this modelling approach. This limitation can be avoided if a significant amount of the parent $\gamma$ is retained at room temperature [19], but this is not the case in the typical C-Mn bainitic steels used in the present study. Another group of "indirect" methods is based on the use of model alloys (with the same stacking fault energy as the analysed steel) that retain their austenitic structure down to room temperature (e.g., [15,20-21]). However, this method requires the complex metallurgical phenomena during the TMCP hot rolling to be reproduced in the model alloy. In particular, as many phenomena are thermally activated (e.g., recrystallisation, grain growth, precipitation), this is not always possible in a reliable way. Consequently, there is a need for a more unambiguous observation of both the $\gamma$ microstructure and the phase transformation characteristics. In this respect, crystallographic approaches for recovering the $\gamma$ microstructure based on $\alpha$ orientation maps measured by EBSD offer the possibility of directly inferring the parent $\gamma$ microstructure from the inherited one [22-26]. The essence of these methods is based on the crystallographic correspondence between the parent grain and the bcc-type product orientation (martensite or bainite). When inspecting an EBSD map of the product structure, the use of these crystallographic ORs determines whether neighbouring product orientations have nucleated from one unique parent grain. The term "neighbouring" means that topological information is needed when recovering the parent phases, i.e., $\alpha$ grains must grow within the $\gamma$ grain. This is only the case for the transformations with a displacive component, as the disciplined motion of atoms cannot persist through changes in crystallographic orientations [27], and as such, the growing products are necessarily restricted to the same parent grain. The basic principles of the $\gamma$ reconstruction process as well as the advantages of each approach have been reviewed by Germain et al. [28]. The reconstruction can be based on the use of average orientations within inherited grains [26] or the local orientation of every pixel [25]. Although the use of averaged orientations of inherited grains reduces computation time, a reconstruction on a local scale (i.e., pixel-based) is preferable because it allows the orientation gradient within $\alpha$ grains to be taken into account in the reconstruction process in highly deformed steels. On the other hand, the parent orientation determination in [25] is based on a one-step minimisation of the angular deviation between the experimental $\alpha$ orientations and the predicted ones, whereas the approach in [26] relies on a robust two-step algorithm. The first step in the latter procedure consists of an unambiguous determination of the parent orientations - or nuclei - with a low tolerance. In a second step, these nuclei are spread to form the complete grains. This methodology has been successfully applied to two low-carbon steels. A combination of the strong points of these two existing models, i.e., the local pixel-by-pixel analysis proposed by Miyamoto et al. [25] and the nuclei identification and spreading strategy proposed by

Germain et al. [26], is expected to give rise to a more accurate $\gamma$ reconstruction. This is the basis of the alternative approach developed in the present study and presented in section 3.1.

The influence of process parameters on the reconstruction quality has already been thoroughly discussed [26,28]. In contrast, there is, to our knowledge, still no experimental evidence to prove the reliability of crystallographic reconstruction. Our focus is therefore on the experimental validation of the proposed reconstruction procedure: first, on the determination of the OR, and second, on the accuracy of the reconstructed prior $\gamma$ grain boundaries. This experimental study is given in section 3.2.

As noted above, the TMCP route can be tailored based on the improved knowledge of the effects of various rolling parameters on both the $\gamma$ conditioning and the $\gamma$-to-$\alpha$ phase transformation. The finish rolling temperature (FRT) is one of the main parameters that affects the $\gamma$ condition prior to transformation, as it determines the lowest temperature at which the deformation is applied. Ray et al. [29] reported a sharp transformation texture in plain C and Nb microalloyed steels for low FRTs, typically in the $\gamma$ non-recrystallisation range. It promotes the {332}<-1-13> component in particular, which is believed to be the most beneficial component for mechanical properties. As this component is mostly inherited from the parent brass {110}<1-12> and S {213}<-3-64> grains, the TMCP route could be designed to strengthen the two latter orientations in $\gamma$. However, the development of a transformation texture is caused by both the deformation texture in $\gamma$ and the variant selection (VS) during the phase transformation. For the reasons expressed above, the experimental investigation of these two separate effects is not direct, and their respective contributions to the transformation texture remain unclear. The deformation of $\gamma$ has been reported to either enhance the onset of bainitic transformation [30] or reduce the onset and kinetics [31]. Shirzadi et al. [32] investigated the influence of ausforming at 600 °C on the bainite orientation and found no strong VS. Gong et al. [33] confirmed the latter result but found that ausforming at a lower temperature (300 °C) accelerates bainite transformation and produces a characteristic microstructure. However, the latter results were obtained for nanobainite steels, which contain more than 1.5 wt% silicon, deformed under uniaxial compression (due to the use of cylindrical specimens). To our knowledge, such advanced analysis of the correlation between the ausforming temperature and the phase transformation behaviour is absent for industrially relevant high-strength steels deformed under plane-strain compression. Therefore, the purpose of this study is to clarify the effect of the FRT on the bainite transformation in hot-rolled C-Mn steels using the alternative crystallographic $\gamma$ reconstruction method developed in the present study. The main question this study explores is: What is the origin of the sharp transformation texture obtained in high-strength C-Mn steels finish rolled at low temperatures?

**2. Experimental**

The material used for the study is a Fe-0.15C-1.5Mn-1.0(Cr+Mo) (wt.%) steel with B addition. Prior to hot rolling, the steel was reheated to 1250°C, and after rough rolling to a thickness of 35 mm, the finish rolling consisted of seven passes to a final thickness of 4 mm. The finish rolling temperature was 829°C, 861°C or 908°C for samples 1, 2 and 3, respectively. The coiling temperature was 430°C for all the materials. These temperatures were measured using a pyrometer. Tensile tests on these materials were performed using standard samples according to EN6892-1. The mechanical properties of these three samples are given in Table 1.

|          | FRT (°C) | YS (MPa) | TS (MPa) | YS/TS | A (%) |
|----------|----------|----------|----------|-------|-------|
| sample 1 | 829      | 1010     | 1120     | 0.90  | 11.5  |
| sample 2 | 861      | 961      | 1121     | 0.86  | 11.6  |
| sample 3 | 908      | 904      | 1048     | 0.86  | 11.7  |

**Table 1.** Measured yield strength (YS), tensile strength (TS) and percent elongation (A) for the analysed samples.

A Bahr DIL805 dilatometer was used to construct the continuous cooling transformation (CCT) diagram of the steel. The samples were reheated to 900°C for 5 min prior to continuous cooling to room temperature. All the temperatures were measured and controlled using a K-type thermocouple welded to the sample. The CCT diagrams shown in Fig. 1 were constructed without deformation and with a 0.3 compressive strain at 900°C. The martensite start (Ms) temperature was estimated as 425°C. As seen from Fig. 1, the martensitic hardenability was suppressed by the $\gamma$ deformation, leading to enhanced bainite formation. The cooling rate after the hot rolling of the steels was chosen to be sufficiently high to avoid the formation of bainite at higher temperatures and allow the formation of bainite just above the Ms temperature, as illustrated in the CCT diagram in Fig. 1. Shortly after the end of the water cooling and the onset of bainitic transformation, the materials were coiled and cooled to room temperature over approximately 24 h. The samples for microstructural examination were taken from the RD-ND cross sections. After polishing, the samples were etched with Picral + $Na_2S_2O_5$.

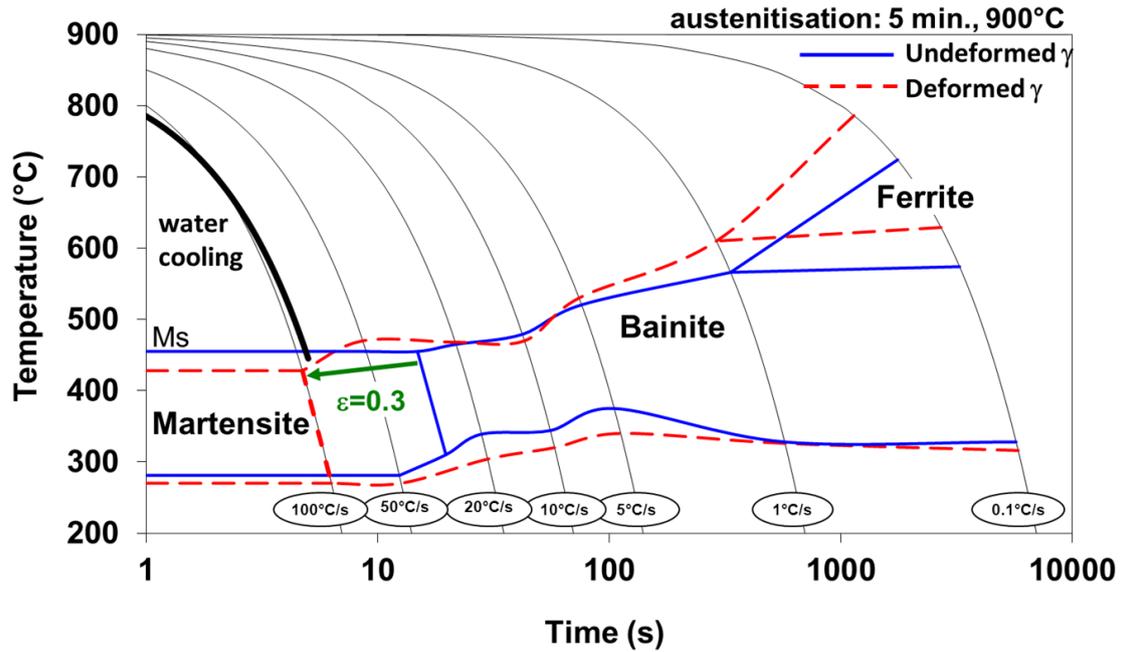

**Fig. 1.** CCT diagram of the undeformed (solid blue) and deformed (dashed red) steel.

Non-ausformed (i.e. without hot rolling) martensite was also formed by direct quenching after austenitizing. This sample, termed the non-deformed sample, is used in section 3.2 in particular to assess the accuracy of the $\gamma$ reconstruction method. In addition, the relevance of the OR refinement procedure on a quenching and partitioning (Q&P) steel is examined. The details of the production of this steel are given elsewhere [34].

For the scanning electron microscopy (SEM) imaging, EBSD and X-ray diffraction (XRD) analyses, samples were prepared using standard grinding and polishing procedures. The mechanically polished surface was etched in Nital-2% concentrated $HNO_3$ in ethanol for a few seconds for SEM, whereas it was polished in a colloidal silica suspension for EBSD experiments. The SEM used in the present study is a field emission gun (FEG) JEOL JSM-7001F equipped with a HKL Nordlys camera for the acquisition of electron backscatter patterns. EBSD mappings were recorded at an accelerating voltage of 20 kV, a working distance of 20 mm and a step size of 0.2 μm. XRD texture measurements were performed on a Bruker D8 diffractometer equipped with a cobalt source. Pole figures for the (110), (310), (211) and (200) reflections were used to calculate the orientation distribution functions (ODFs) using a truncation limit of 22.

## 3. Reconstructing the prior austenite microstructure

3.1. Principle

The present reconstruction procedure is carried out in three steps:

*(i) Orientation relationship refinement*

The material after transformation acquires a texture that is related in a fairly precise way to the texture of the parent $\gamma$. The first step is thus to determine the crystallographic OR that describes the transformation. The Kurdjumov-Sachs (KS) (i.e., $\{111\}_\gamma //\{110\}_\alpha$, $<110>_\gamma //<111>_\alpha$) [35] or Nishiyama-Wassermann (NW) (i.e., $\{111\}_\gamma //\{110\}_\alpha$, $<211>_\gamma //<110>_\alpha$) [36] are the most frequently employed ORs. Note that KS variants share parallel close-packed planes (CPs) and parallel close-packed direction (CDs) with the parent grain. However, ORs intermediate between KS and NW are usually measured in steels from EBSD and TEM experimental studies [e.g., 37-39].

Therefore, the OR must be experimentally determined for each tested specimen, and two separate methods have been proposed by Miyamoto et al. [40] and Humbert et al. [41] for this purpose. Although the mathematical approach is different for each, both rely only on the measured $\alpha$ variant orientations, without the need of the retained $\gamma$ in the inherited structure. The approach described in [40] is used in the present reconstruction procedure. It consists of minimising, on a number $N$ of $\alpha$ orientations transformed from the same $\gamma$ grain, the average deviation angle $\overline{\Delta\theta}$ between $\alpha$ orientations theoretically predicted by the OR and experimental $\alpha$ orientations:

$$\overline{\Delta\theta} = \sum_N arccos\{(D_i[1,1] + D_i[2,2] + D_i[3,3] - 1)/2\}/N \qquad (1)$$

where $D_i[1,1]$, $D_i[2,2]$ and $D_i[3,3]$ are diagonal components of the deviation matrix $D_i$ of the $i$ th data point, defined as

$$D_i = \Delta g \left( C_{i,j} M^\gamma \right) \left( C_{i,k} M_i^\alpha \right)^{-1} \qquad (2)$$

with $\Delta g$ is the OR, $M^\gamma$ is the gamma orientation, $M_i^\alpha$ is the measured $\alpha$ orientation of the $i$ th data point, and $C_{i,j}$ and $C_{i,k}$ are one of the 24 symmetry operators in the cubic system for the $i$ th data point (chosen such that the misorientation angle of $D_i$ is minimal). The $\overline{\Delta\theta}$ minimisation is performed by numerical fitting with $M^\gamma$ and $\Delta g$ as independent variables. The reader is referred to the work of Miyamoto et al. [40] for more details. Note that an averaged OR is derived from this approach, i.e., any possible dispersion of OR's during the phase transformation is not considered. A discussion on the physical relevance of the averaged OR is proposed in [42].

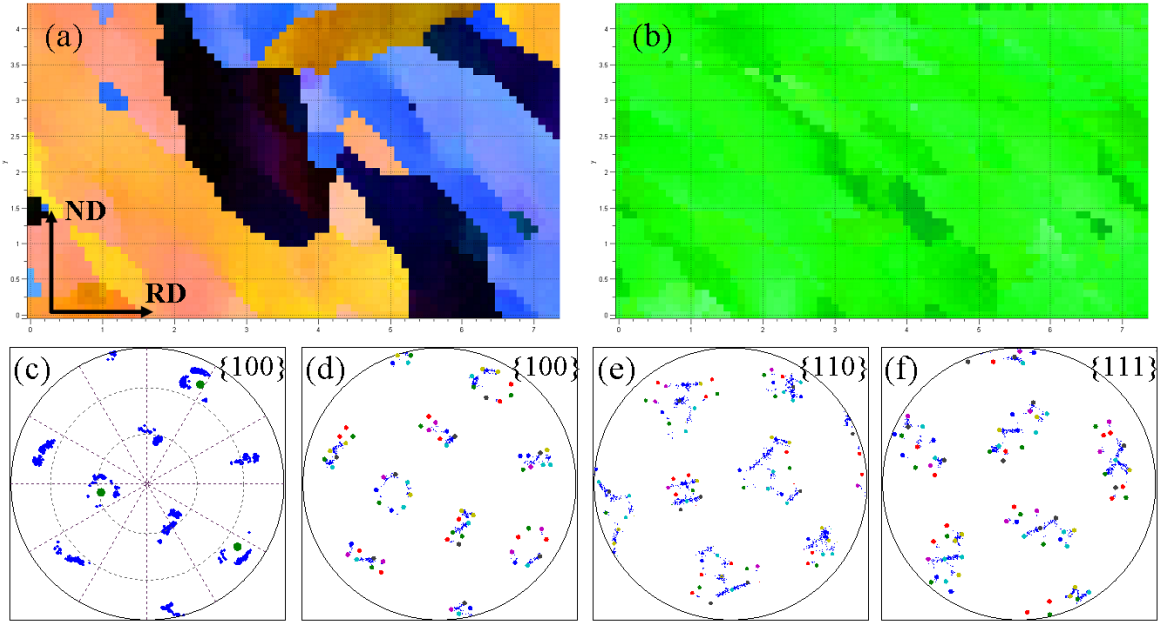

**Fig. 2.** (a) Experimental $\alpha$ and (b) reconstructed $\gamma$ orientation mappings along the RD. The colour coding is given in Fig. 3(a). Stereographic projections of the {100} reconstructed $\gamma$ orientations (green dots) and (d-f) the {100}, {110} and {111} predicted $\alpha$ orientations (coloured dots). The stereographic projection of experimental $\alpha$ (blue dots) is superimposed in (c-f).

Fig. 2(a) shows an inverse pole figure (IPF) mapping along the rolling direction (RD) of a region of $\alpha$ orientations transformed from only one $\gamma$ grain in the non-deformed sample (colour legend in the inset of Fig. 3(a)). The results show an OR with angles between CPs and CDs of approximately 1.4° and 2.5°, respectively. This OR is determined after the fitting is completed at $\overline{\Delta\theta}$ =2.3°. Note that this value does not correspond to the OR determination accuracy, as it includes (in addition to the EBSD experimental errors) the $\alpha$ orientation spread due to the $\gamma$ misorientation issued from both the hot-rolling process and the phase-transformation-induced strain. A spreading of ORs may also occur. Miyamoto et al. [40] proved that $\Delta g$ can be determined within an error of 1°. The stereographic projection in Fig. 2(c) of the measured <100>$_\alpha$ (from Fig. 2(a)) and reconstructed <100>$_\gamma$ (from Fig. 2(b)) directions are represented by blue and green dots, respectively. The <100>$_\alpha$ projections form three Bain zones, each being composed of one truncated square and two bars. It can be verified in Fig. 2(c) that the <100>$_\gamma$ poles of prior $\gamma$ correspond to the centres of truncated squares. The {100}, {110} and {111} pole figures in Figs. 2(d-f) of the measured $\alpha$ orientations (blue dots) and predicted $\alpha$ orientations using the fitted couple "$\gamma$ orientation – OR" (coloured dots) indicate the good agreement between the experimental and calculated data.

*(ii) Identification of $\gamma$ nuclei*

This step consists of determining whether the $\alpha$ orientations in each cropped region of the EBSD mapping are transformed from only one $\gamma$ grain; if so, the analysed region is referred to as a $\gamma$ nucleus. For such nucleus to exist, two criteria must be fulfilled.

First, the 24 potential parents $M_{i,l}^{\gamma}$ are calculated for all the $i$ th $\alpha$ orientations represented in the cropped region using the OR $\Delta g$ measured in step *(i)*:

$$M_{i,l}^{\gamma} = C_{i,j}^{-1} \Delta g^{-1} C_{i,k} M_i^{\alpha} \tag{3}$$

with the subscript $l$ ranging from 1 to 24 describing the 24 symmetrically equivalent parent orientations. Next, the two most intense orientations, $M_1^{\gamma}$ and $M_2^{\gamma}$, of this set of potential parents in the cropped region are calculated using the free and open-source MTEX toolbox [43] running in MATLAB (MathWorks, Inc., Natick, MA, USA). The modal values of these two orientations are denoted $f_1^{\gamma}$ and $f_2^{\gamma}$, with $f_1^{\gamma} \geq f_2^{\gamma}$ by convention. If the intensity ratio $f = (f_1^{\gamma} - f_2^{\gamma})/f_1^{\gamma}$ is larger than a predefined threshold $f_t$, the parent orientation $M_1^{\gamma}$ is considered a unique parent orientation, and the first criterion is met.

Second, $\overline{\Delta\theta}$ is calculated for the cropped region using Eqs. (1-2) with $\Delta g$ measured in step *(i)* and $M_1^{\gamma}$ measured in the $\gamma$ orientation. The second criterion is met when $\overline{\Delta\theta}$ is lower than a predefined threshold $\overline{\Delta\theta}_t$. If both criteria are fulfilled, the analysed variants are then assumed to be inherited from the same austenitic grain. It should be noted that a $\overline{\Delta\theta}$ minimisation by numerical fitting with $M_1^{\gamma}$ as variable could be *a priori* useful for obtaining more accurate nuclei detection. However, it has been experimentally checked that such a fitting results in a small 5-10% decrease of $\overline{\Delta\theta}$, and of special interest, it does not modify the couple $\{C_{i,j}, C_{i,k}\}$, which is used to derive the parent orientation of every data point.

*(iii) Spreading of $\gamma$ nuclei*

The $\gamma$ nuclei cannot cover the complete EBSD mapping, mostly due to the topological differences between the square cropped regions and the irregular shape of the prior $\gamma$ grains. Therefore, the pixels to which no parent orientation is assigned still remain after the nuclei identification step. A pixel-by-pixel analysis is then carried out to determine whether a non-assigned data point can be incorporated into an existing $\gamma$ nucleus. This analysis is described below. Consider that the $i$ th data point is non-assigned, while at least one of its eight neighbours, represented by the subscript $p$ below, belongs to a $\gamma$ nucleus (i.e., inherited from a $\gamma$ grain whose orientation is given by $M_p^\gamma$). As previously performed, the 24 potential parents $M_{i,l}^\gamma$ for the $i$ th $\alpha$ orientation are first calculated using Eq. (3). Next, the deviation matrix $\Omega_{i,l}^p$ between the $i$ th potential parents and the parent orientation for all assigned $p$ th neighbours is obtained as follows:

$$\Omega_{i,l}^p = M_p^\gamma \left( M_{i,l}^\gamma \right)^{-1} \tag{4}$$

The misorientation angle $\varpi_{i,l}^p$ and its minimum value $\varpi_i^p$ are then derived:

$$\varpi_i^p = min(\varpi_{i,l}^p), \text{ with } \varpi_{i,l}^p = arccos\{(\Omega_{i,l}^p[1,1] + \Omega_{i,l}^p[2,2] + \Omega_{i,l}^p[3,3] - 1)/2\} \tag{5}$$

If $\varpi_i^p$ is lower than a predefined threshold $\varpi_t$, the $i$ th $\alpha$ orientation is incorporated into the $\gamma$ nucleus containing the $p$ th neighbour. In practice, this step is incrementally performed with increasing $\varpi_t$ values.

The reconstruction quality was optimised through a trial-and-error process that adjusts the above-mentioned thresholds. The results show that $f_t = 0.2$, $\overline{\Delta\theta}_t = 3°$ and $\varpi_t$ from 3° to 10° yield the best results. The size of cropped regions must be adjusted for every EBSD mapping as a function of the inherited grain size: the larger the cropped area, the more accurate the parent orientation determination (due to the higher number of inherited variants) but also the poorer the spatial resolution of the reconstructed map; 3x3 µm$^2$ squares with 0.6 µm between neighbouring squares are used in the present work.

3.2. Experimental validation

The use of retained $\gamma$ in the inherited structure is appropriate to investigate the relevance of the refined OR, assuming that the "stable" $\gamma$ has not rotated upon cooling. Therefore, we use a Q&P steel grade with ~ 10 wt% retained $\gamma$. The IPF mapping along the RD is shown in Fig. 3(a). The phase map is given in Fig. 3(b) with bcc and fcc phases represented by blue and red areas, respectively. The comparison between the measured and refined ORs must be conducted on an $\alpha$ region inherited from only one $\gamma$ grain showing no annealing twins; therefore, the first step aims to select such an $\alpha$ region from the EBSD mapping.

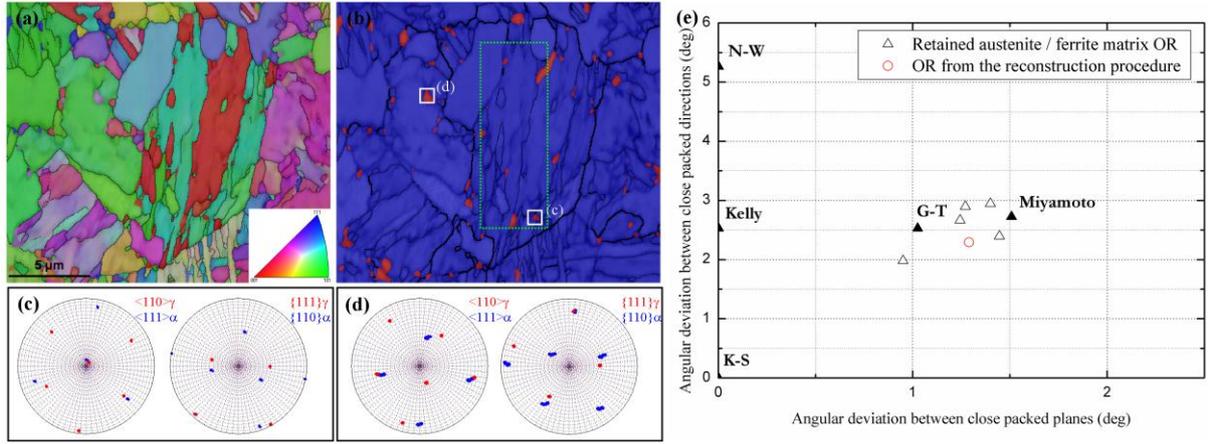

**Fig. 3.** (a) Experimental $\alpha$ orientation mapping along the RD on the Q&P steel together with the colour coding (inset), (b) associated phase mapping with bcc and fcc in red and blue, respectively. (c-d) Stereographic projections of the $\gamma$ and $\alpha$ CPs and CDs from the areas marked by the white rectangles in (b). (e) Comparison of the OR measured from the RA particles and the refined OR from the dotted green rectangle shown in (b).

One strategy relies on the misorientation angles between variants inherited from the same $\gamma$ grain. As seen from Table 2, all K-S variants are misoriented with respect to each other by either less than 21.1° or greater than 47.1° (similar results are obtained for refined ORs, which are only slightly deviated from the K-S variant; see Table 2).

| No. | Austenite plane/direction | Ferrite plane/direction | Rotation axis | B-H slip system | CP group | Bain group | Misorientation angle/axis from V1 (K-S OR) | Misorientation angle from V1 (sample 1 OR) | Misorientation angle from V1 (sample 2 OR) | Misorientation angle from V1 (sample 3 OR) | Closest variant using refined OR's |
|---|---|---|---|---|---|---|---|---|---|---|---|
| 1 | (1-1-1)[-1-10] | (-1-10)[-111] | [1-12] | -cIII | CP3 | B1 | - | - | - | - | 11 |
| 2 | (1-1-1)[110] | (110)[-111] | [-1-12] | cIII | CP3 | B2 | 60/[0.577,-0.577,-0.577] | 60.3 | 60.2 | 60.1 | 21 |
| 3 | (-1-1-1)[-1-10] | (-1-10)[1-11] | [1-1-2] | dIII | CP4 | B2 | 49.5/[0.577,-0.577,0.577] | 53.1 | 51.7 | 49.4 | 17 |
| 4 | (-1-1-1)[110] | (110)[1-11] | [-112] | -dIII | CP4 | B1 | 10.5/[0.577,-0.577,0.577] | 9.2 | 8.8 | 10.8 | 13 |
| 5 | (11-1)[1-10] | (1-10)[-1-11] | [112] | bIII | CP2 | B2 | 57.2/[-0.738,0.628,-0.246] | 56.7 | 57.5 | 57.1 | 19 |
| 6 | (11-1)[-110] | (-110)[-1-11] | [-1-1-2] | -bIII | CP2 | B1 | 20.6/[-0.659,0.363,-0.659] | 16.8 | 15.0 | 16.9 | 9 |
| 7 | (-1-1-1)[1-10] | (1-10)[111] | [11-2] | -aIII | CP1 | B1 | 21.1/[0.410,0,-0.912] | 19.2 | 16.4 | 17.5 | 15 |
| 8 | (-1-1-1)[-110] | (-110)[111] | [-1-12] | aIII | CP1 | B2 | 57.2/[-0.628,-0.246,-0.738] | 56.7 | 57.5 | 57.1 | 23 |
| 9 | (11-1)[-10-1] | (-10-1)[-1-11] | [1-2-1] | bII | CP2 | B1 | 14.9/[-0.354,-0.933,-0.065] | 13.6 | 11.6 | 12.4 | 6 |
| 10 | (11-1)[101] | (101)[-1-11] | [-121] | -bII | CP2 | B3 | 50.5/[-0.490,-0.739,0.463] | 50.7 | 52.6 | 52.1 | 20 |
| 11 | (1-1-1)[-10-1] | (-10-1)[-111] | [-1-21] | cII | CP3 | B1 | 10.5/[-0.707,0,-0.707] | 5.4 | 6.2 | 8.5 | 1 |
| 12 | (1-1-1)[101] | (101)[-111] | [12-1] | -cII | CP3 | B3 | 60/[-0.707,0,-0.707] | 57.9 | 60.1 | 59.1 | 22 |
| 13 | (-1-1-1)[10-1] | (10-1)[1-11] | [121] | dII | CP4 | B1 | 14.9/[0.933,0.354,-0.065] | 13.6 | 11.6 | 12.4 | 4 |
| 14 | (-1-1-1)[-101] | (-101)[1-11] | [-1-2-1] | -dII | CP4 | B3 | 57.2/[0.603,-0.357,0.714] | 55.9 | 59.5 | 59.3 | 18 |
| 15 | (-1-1-1)[10-1] | (10-1)[111] | [-1-21] | aII | CP1 | B1 | 20.6/[0,0.955,0.296] | 18.4 | 16.2 | 18.1 | 7 |
| 16 | (-1-1-1)[-101] | (-101)[111] | [1-21] | -aII | CP1 | B3 | 47.1/[0.626,0.719,0.302] | 52.8 | 51.5 | 49.1 | 24 |
| 17 | (-11-1)[0-1-1] | (0-1-1)[1-11] | [21-1] | -dI | CP4 | B2 | 50.5/[-0.490,-0.463,0.739] | 50.7 | 52.6 | 52.1 | 3 |
| 18 | (-1-1-1)[011] | (011)[1-11] | [-2-11] | dI | CP4 | B3 | 50.5/[0.186,-0.615,-0.767] | 51.7 | 53.4 | 53.1 | 14 |
| 19 | (11-1)[0-1-1] | (0-1-1)[-1-11] | [2-11] | -bI | CP2 | B2 | 47.1/[-0.302,0.626,0.719] | 52.8 | 51.5 | 49.1 | 5 |
| 20 | (11-1)[011] | (011)[-1-11] | [-2-1-1] | bI | CP2 | B3 | 51.7/[0.659,-0.659,0.363] | 48.7 | 53.4 | 54.5 | 10 |
| 21 | (1-1-1)[01-1] | (01-1)[-111] | [-2-1-1] | cI | CP3 | B2 | 60/[0,0.707,-0.707] | 57.9 | 60.1 | 59.1 | 2 |
| 22 | (1-1-1)[0-11] | (0-11)[-111] | [211] | -cI | CP3 | B3 | 49.5/[0,0.707,-0.707] | 55.9 | 54.0 | 51.9 | 12 |
| 23 | (-1-1-1)[01-1] | (01-1)[111] | [-211] | -aI | CP1 | B2 | 57.2/[0.357,-0.603,0.714] | 55.9 | 59.5 | 59.3 | 8 |
| 24 | (-1-1-1)[0-11] | (0-11)[111] | [2-1-1] | aI | CP1 | B3 | 50.5/[-0.615,0.767,0.186] | 51.7 | 53.4 | 53.1 | 16 |

**Table 2.** Description of the 24 variants considered in the present study. The rotation axis is given by the cross product of the respective Burgers vector and the {111} slip plane normal. CP classification: CP1, (111)$\gamma$ //(011)$\alpha$; CP2, (-1-11)$\gamma$ //(011)$\alpha$; CP3, (-111)$\gamma$ //(011)$\alpha$; CP4, (1-11)$\gamma$ //(011)$\alpha$. Bain classification: B1, (100)$\gamma$ //(100)$\alpha$, [010]$\gamma$ //[011]$\alpha$; B2, (010)$\gamma$ //(010)$\alpha$, [001]$\gamma$ //[101]$\alpha$; B3, (001)$\gamma$ //(001)$\alpha$, [100]$\gamma$ //[110]$\alpha$.

In other words, two neighbouring variants disoriented by an angle between 21.1° and 47.1° are likely inherited from different $\gamma$ grains, and consequently, a prior $\gamma$ grain boundary (GB) can be placed between these two variants. Note that the opposite is not true; that is, a prior $\gamma$ GB is not necessarily located between variants disoriented between 21.1° and 47.1°.

A fraction of the prior $\gamma$ GBs in the Q&P sample is therefore revealed by the GBs between 21.1° and 47.1° drawn in bold black lines in Fig. 3(b). The OR refinement method is applied to the region defined by the dashed green rectangle containing *a priori* no such boundary. The calculated OR is indicated by the open red circle in Fig. 3(e), together with the well-known K-S, N-W, Kelly [44], Greninger-Troian [45] and Miyamoto [40] ORs. In addition, the orientations of five retained $\gamma$ particles (located in the dashed green rectangle) have been measured with respect to their respective surrounding matrix orientation. Stereographic projections of CPs and CDs for two of these $\alpha$ and $\gamma$ orientations are shown in Figs. 3(c-d). As seen from Fig. 3(e), the five measured ORs (indicated by open black triangles) deviate from the calculated OR by less than the 1° uncertainty reported earlier by Miyamoto et al. [40], which confirms experimentally the relevance of the OR refinement method. The spread of the measured ORs is interpreted to represent the inheritance of the retained $\gamma$ orientation from the plastically accommodated prior $\gamma$ structure. The obtained results have been validated on other prior $\gamma$ grains. Note that the stereographic projection in Fig. 3(d) shows that the OR does not vary considerably over the EBSD mapping of the Q&P steel.

A comparison between the reconstructed $\gamma$ mapping and the microstructure revealed by Béchet-Beaujard (BB) etching [10] is performed to examine the accuracy of the reconstructed prior $\gamma$ grain boundaries. To ensure maximum efficiency of the BB etching, the comparison is conducted on the quenched steel grade, which is referred to as the 'non-deformed sample'. Fig. 4(a) shows the IPF mapping along the RD acquired on this steel and provides evidence of a fully lath martensite microstructure. The reconstructed $\gamma$ mapping is given in Fig. 4(b), while ODFs at $\varphi_2=45°$ (following Bunge notation [46]) are given in the inset in the latter figures. For an accurate comparison to be made, two detailed views are visualised in Fig. 5: the reconstructed $\gamma$ mapping in (a) and (d), the experimental $\alpha$ mapping with GBs between 21.1° and 47.1° in bold black lines in (b) and (e), and the optical microscope images of the BB etched sample in (c) and (f).

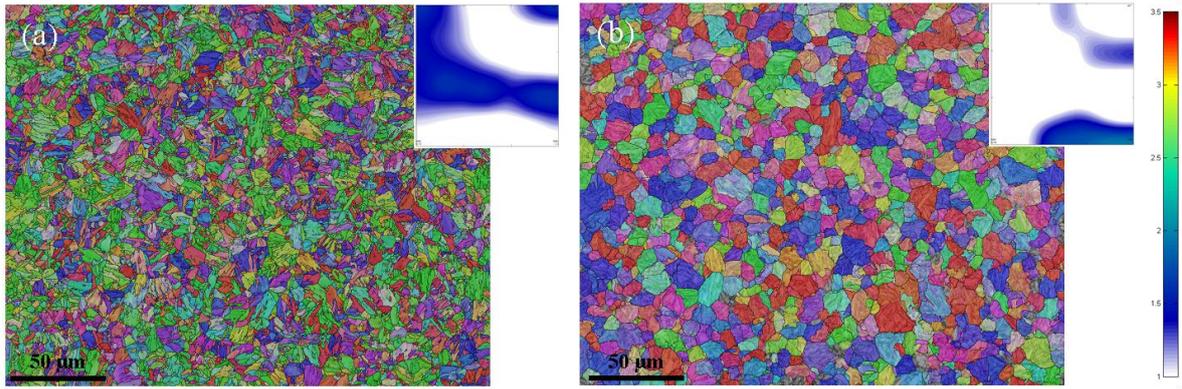

**Fig. 4.** (a) Experimental $\alpha$ and (b) reconstructed $\gamma$ orientation mapping along the RD together with the ODF sections at $\varphi_2 = 45°$ (inset). The colour coding is given in Fig. 3(a).

As seen from Fig. 5, the reconstructed prior $\gamma$ boundaries appear to be in good overall agreement with those revealed by BB etching. Three observations must be stressed here. First, a few reconstructed $\gamma$ grains exhibit questionable boundaries (dashed orange arrows), mostly due to the irregular, unusual grain shape and the absence of grain boundaries between 21.1° and 47.1° in the $\alpha$ mapping (note that BB etching is particularly inefficient in zones containing these dubious boundaries). Second, the reconstruction process offers greater added value with respect to the 21.1°-47.1° boundary angles strategy, which locally reveals only a part of the $\gamma$ grain boundaries (dashed red arrows); these unclosed boundaries are successfully reconstructed by the present procedure and are confirmed to be present in the sample after BB etching. Third, the reconstructed $\gamma$ annealing twins (red boundaries in the $\gamma$ mappings using the Brandon criterion [47]) are not observed after BB etching (solid arrows), as previously reported [26], most likely due to a low amount of segregated elements such as phosphorous in the highly ordered $\Sigma 3$ boundary structure. We checked that these reconstructed twins are coherent on a few grains.

However, the reconstructed twin boundaries follow jagged contours instead of ordinary straight lines. This artefact, clearly due to the cropped squares used in the reconstructed mapping, may be avoided after a semi-automated completion, as carried out in [26]. However, this additional step is not performed in the present procedure, mainly because it would require a manual grain-by-grain examination and therefore a longer processing time involving the user. In addition, Miyamoto et al. [25] proved that the use of refined ORs allows the misindexing of twin orientations to be greatly reduced compared to K–S- or N–W-based reconstructions (due to the mirror symmetry in the stacking of (011) $\alpha$ planes in the latter

ORs). Most of the reconstruction mistakes using a measured OR are confirmed to be related to the shape and the exact location of twin boundaries [25] but do not question their presence in the structure. This finding is supported by the higher fraction of twin boundaries measured in the non-deformed sample than in hot-rolled steels (not shown here). All these results prove that the main output of an additional grain-by-grain analysis would only be a $\Sigma 3$ 'boundary straightening' and thus without any significant effect on the relevant grain features (e.g., size, aspect ratio, orientation) used to characterise the prior austenitic microstructure.

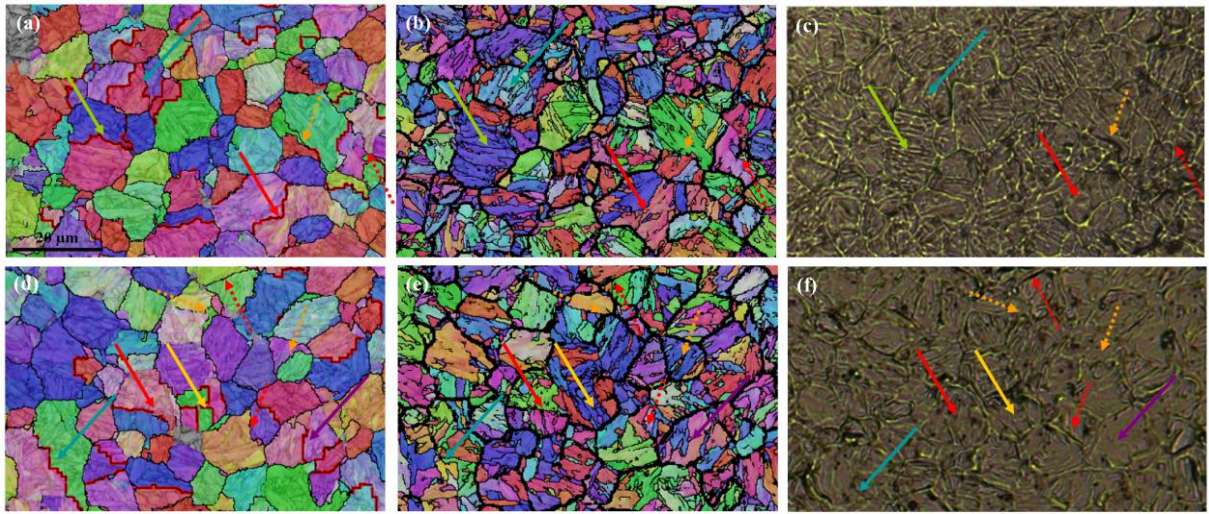

**Fig. 5.** (a), (d) Reconstructed $\gamma$ (GBs > 10° in black); (b), (e) experimental $\alpha$ (GBs between 21 and 47° in black); (c), (f) Béchet-Beaujard etching micrographs. The meaning of the coloured arrows is explained in the text.

## 4. Results

4.1. Microstructure

All the materials essentially consist of bainite with inter- and intralath cementite, as shown in the micrographs of Figs. 6(a-c). This morphology is very similar to what is commonly known as lower bainite for isothermally transformed bainite [48]. For the three rolling conditions, martensite-austenite (M/A) islands are present, as seen in the SEM image in Fig. 6(d). These M/A islands are aligned parallel to the rolling plane and are reported to originate from the local microsegregation of manganese [49]. The morphology of the prior $\gamma$ grains can be distinguished to a limited extent. Especially in the cases of samples 1 and 2, the prior $\gamma$ grains are flattened. Regarding sample 3, rolled at 908°C, the prior $\gamma$ structure is not sufficiently revealed to arrive at a clear conclusion. Experimental IPF mappings along the RD acquired for samples 1, 2 and 3 are shown in Figs. 7 (a-c), together with their respective reconstructed $\gamma$ mappings in Figs. 7 (d-f). Pancake $\gamma$ grains are clearly visualised for samples 1 and 2, in agreement with the above optical images. In contrast, the reconstruction process reveals a near-equiaxed $\gamma$ morphology for sample 3.

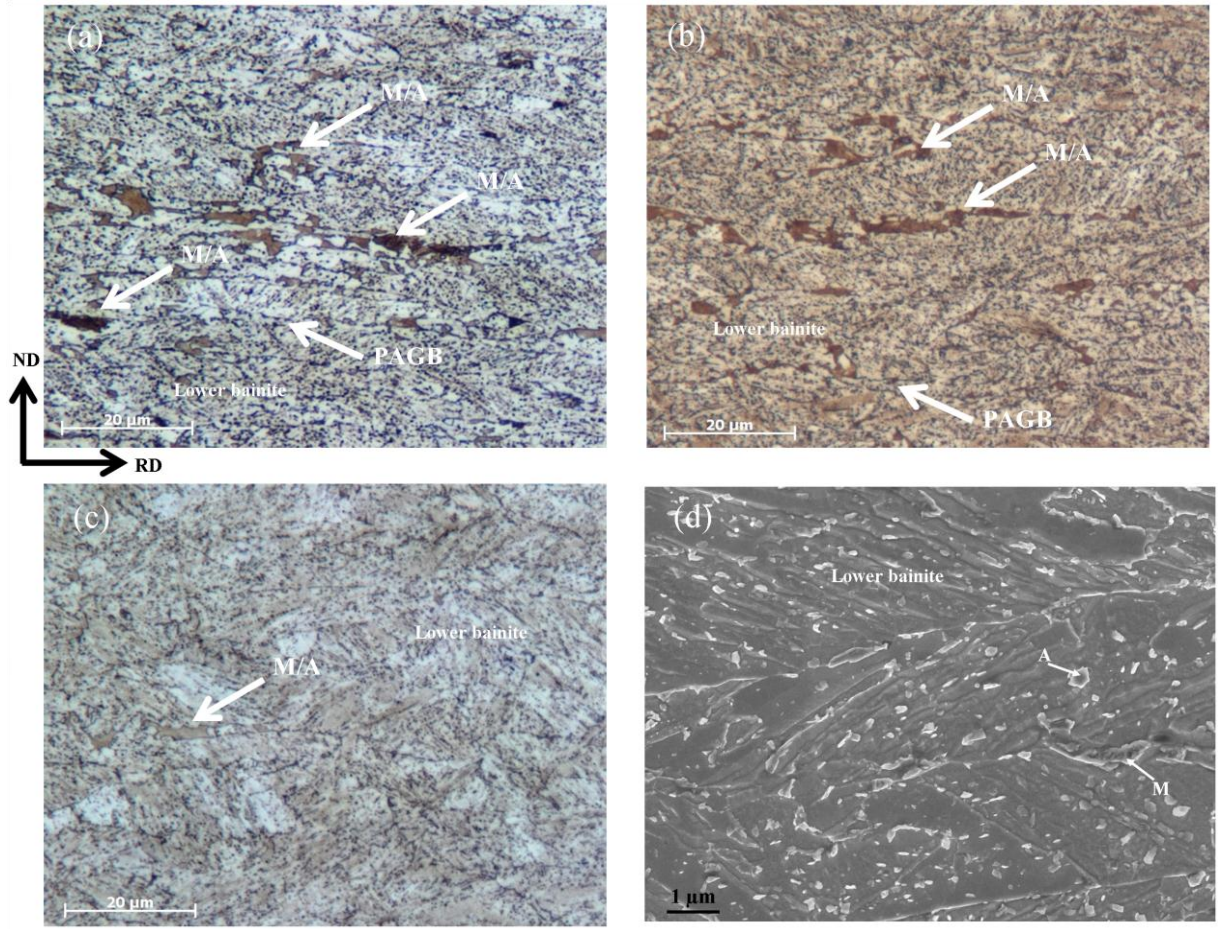

**Fig. 6.** (a-c) Optical micrographs of samples 1, 2 and 3, respectively, after Picral etching. (d) SEM image of sample 1 after Nital etching.

The histograms in Figs. 8(a-e) show the percentage of covered area as a function of various grain parameters. Note that the covered areas are preferred over the frequency distribution profiles to drastically reduce the effect of acquisition artefacts (even though we use a minimum grain size of 10 pixels, as suggested by the ASTM procedure E2627-10). The grain size corresponds here to the equivalent circle diameter. Fig. 8(a) indicates that the $\gamma$ grains are significantly larger in sample 3; the median grain sizes are equal to 10.2, 10.3 and 16.1 µm for samples 1, 2 and 3, respectively. The $\gamma$ grain sizes are fairly similar for samples 2 and 3. Similarly, the grain average misorientations (GAMs) and aspect ratios (ARs) of $\gamma$ grains shown in Figs. 8 (b-c) are similar for samples 1 and 2. In contrast, sample 3 exhibits much lower ARs (linked to the near-equiaxed grain shapes) and GAMs. This less heavily deformed austenitic structure in sample 3 gives rise to larger $\alpha$ grains compared to the other samples (see Fig. 8(d)).

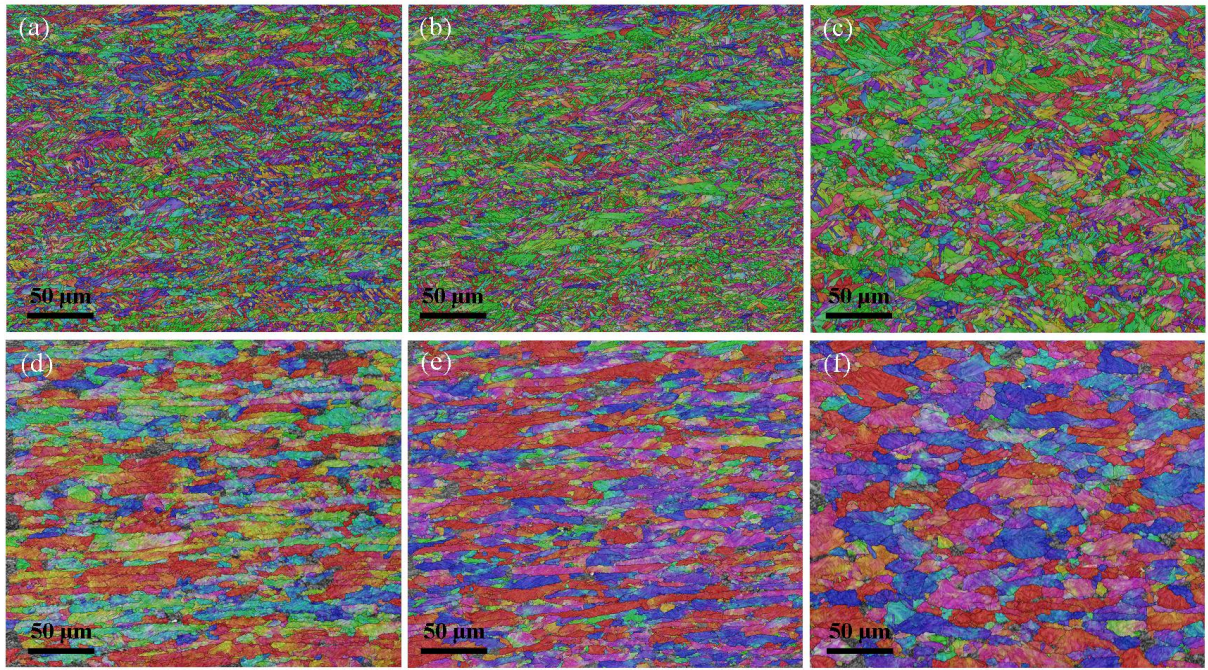

**Fig. 7.** (a-c) Experimental $\alpha$ and (d-f) reconstructed $\gamma$ orientation mappings along the RD of samples 1, 2 and 3, respectively. The colour coding is given in Fig. 3(a).

Furthermore, it is worth considering how the GAM differences among the samples are distributed over grains of different sizes. For this purpose, the GAM distribution is plotted as a function of the grain area using a logarithmic scale to handle the large range of grain areas. For better visualisation, the average GAMs are calculated for equally sized bins of grain area. The plots are shown in Fig. 8(f) for all the samples. The presence of a minimum of 10 grains is verified for every bin to ensure the relevance of the average GAM value. Fig. 8(f) shows that the GAMs have no grain size dependence for the non-deformed steel, i.e. the martensitic reference condition. In contrast, the GAMs of the rolled samples increase linearly with the logarithm of the grain area, with misorientation angles for the largest grains of up to approximately 12°. Interestingly, the GAMs of sample 3 are continuously lower than those of samples 1 and 2 for the complete range of grain areas.

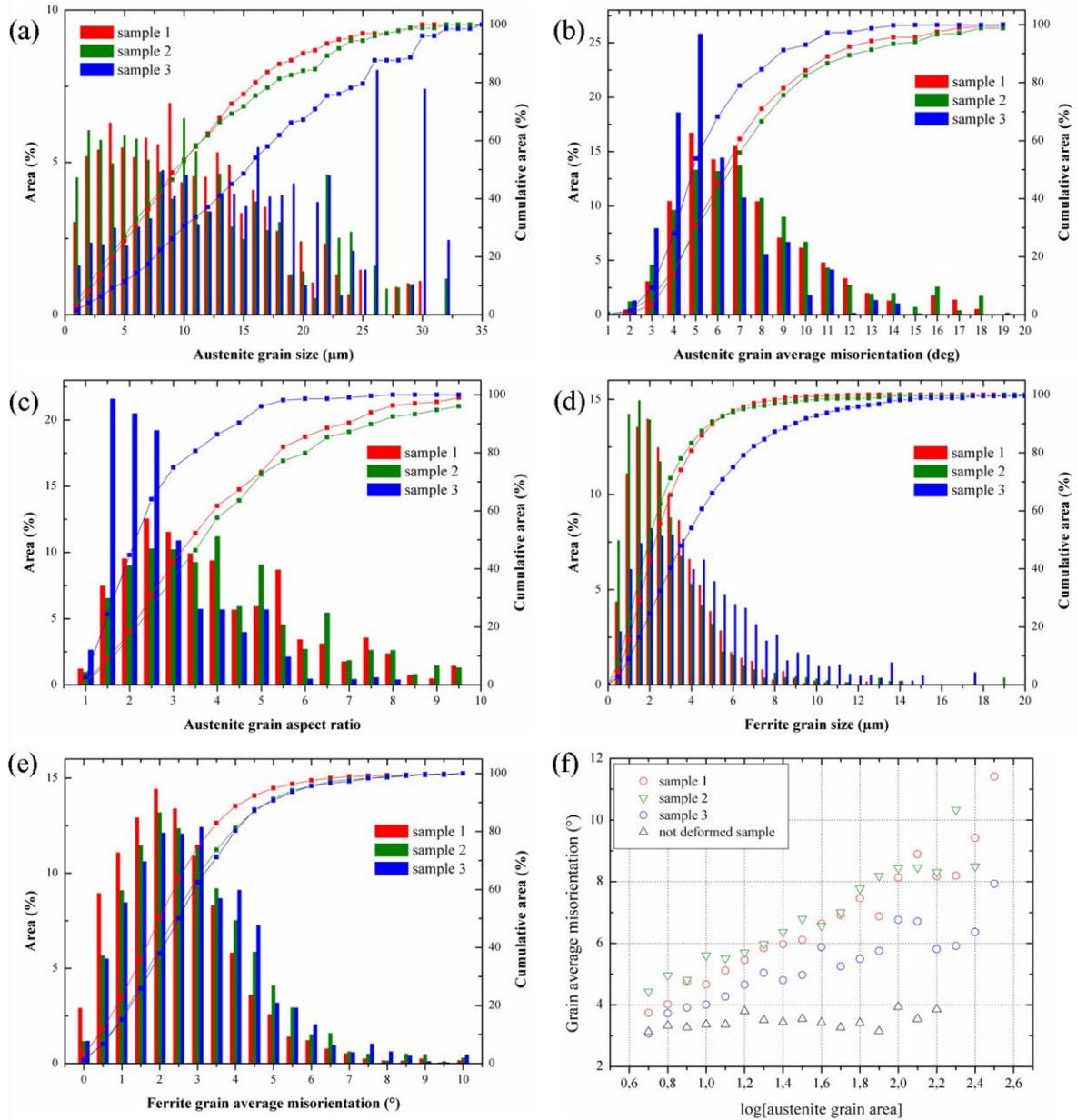

**Fig. 8.** Histograms showing the percentage of covered area as a function of (a) $\gamma$ grain size, (b) $\gamma$ GAM, (c) $\gamma$ AR, (d) $\alpha$ grain size and (e) $\alpha$ GAM. (f) $\gamma$ GAM as a function of the logarithm of the $\gamma$ grain area.

4.2. Orientation relationship

The orientation relationships (ORs) obtained from the refinement procedure (described in section 3.1) are presented in Fig. 9 in terms of angular deviations between $\alpha$ and $\gamma$ CPs and CDs, plotted on the x-axis and y-axis, respectively (note the difference in scale between the two axes). This procedure has been applied to five prior $\gamma$ grains for each sample. The results show that the vast majority of ORs, which are quite close to the G-T [44] and Miyamoto [40]

ORs, exhibit deviations between CDs and CPs of 1.5° to 3.5° and 0.5° to 2.5°, respectively. Of particular interest, the range of angular deviations between CDs is substantially lower when considering the data for each sample separately. Furthermore, it can be noted that the higher the FRT, the lower the angular deviation between CDs. In contrast, the deviations between CPs are more scattered and rather insensitive to the FRT.

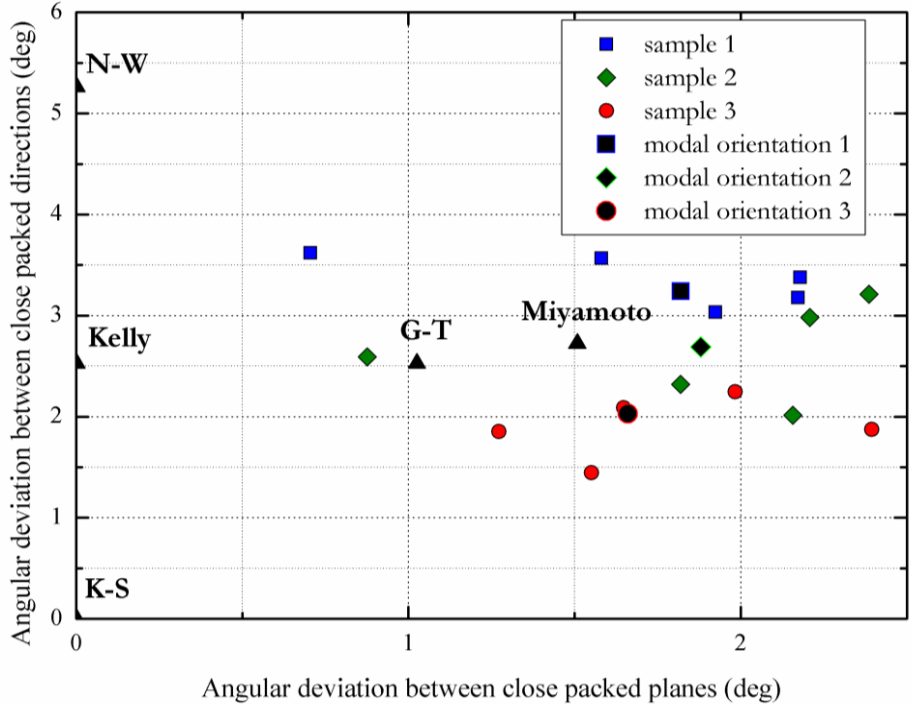

**Fig. 9.** Refined ORs measured for samples 1, 2 and 3.

4.3. Texture

The ODF sections at $\varphi_2 = 45°$ (calculated from the EBSD mappings) for samples 1, 2 and 3 are displayed in Figs. 10(a-c), respectively. An ODF section at $\varphi_2 = 45°$ of the frequently observed texture components in steels is given for reference in Fig. 10(f). The $\alpha$ texture of all the samples is mostly composed of RD//<110> and tilted ND//<111> fibres (i.e., from ~{113}<1-10> to {332}<-1-13> components) together with rotated cube H, H' {100}<110>, rotated Goss L{110}<1-10> and Goss G {110}<001> components. Although the texture component type is similar for all three samples, the texture is sharper in sample 1, with intensities ranging between 2 and 3.5 for the aforementioned components. In addition, the {332}<-1-13> orientation is clearly the most intense one in sample 1, whereas this orientation has approximately the same intensity as the other texture components in samples 2 and 3.

The EBSD mappings shown in Fig. 7 cover a relatively small area (250x200 μm$^2$) compared to the thickness of the steel (4 mm). Therefore, it is crucial to ensure that the texture analysed by EBSD is fairly representative of the whole material. To this end, the texture (averaged over the steel thickness) is also measured by XRD on sample cross-sections using an X-ray beam with a 2-3-mm spot size. The XRD results, shown in Fig. 10(d-e) for samples 1 and 3, respectively (the XRD texture of sample 2 being similar to that of sample 3), confirm that (i) the texture of sample 1 is especially strong, with components up to an intensity level of 6, and (ii) the {332}<-1-13> orientation prevails over the other texture components in sample 1, which is not the case for the other samples. The difference in the intensity of components between XRD and EBSD ODF sections is most likely due to the difference in the measured area. The intensity ratio between the different components is however similar for the two latter techniques.

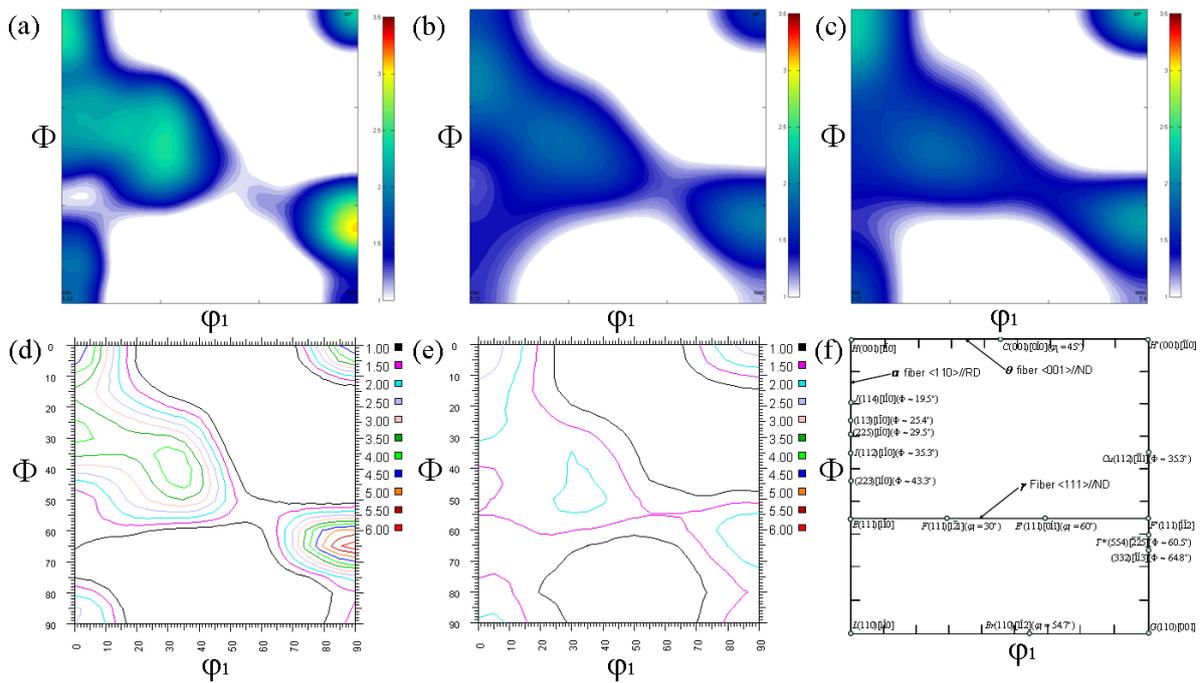

**Fig. 10.** ODF sections at $\varphi_2 = 45°$ of $\alpha$ grains in (a-c) samples 1, 2 and 3, respectively, measured by EBSD; (d-e) samples 1 and 3, respectively, measured by XRD. (f) The frequently observed texture components in steels.

The $\gamma$ texture is derived from the reconstructed mappings illustrated in Figs. 7(d-f). The $\gamma$ ODF sections at $\varphi_2 = 45°$ shown in Figs. 11(a-c) reveal the presence of a typical fcc deformation texture [18,50-51], often described as the β-fibre stretching from the copper {112}<-1-11>, via S {213}<-3-64>, to the brass {110}<1-12> orientations. The intensities of the copper, brass and Goss components are fairly similar for all the samples. In contrast, the fraction of cube-oriented grains is slightly larger in sample 1 than in sample 3. Sample 2 exhibits the lowest cube intensity. The $\gamma$ ODF sections at $\varphi_2 = 63°$ are also shown to evaluate the intensity of the S{213}<-3-64> component, which does not appear in the $\varphi_2 = 45°$ section. As seen from Figs. 11(d-f), the S orientation exhibits the same strength for all three samples.

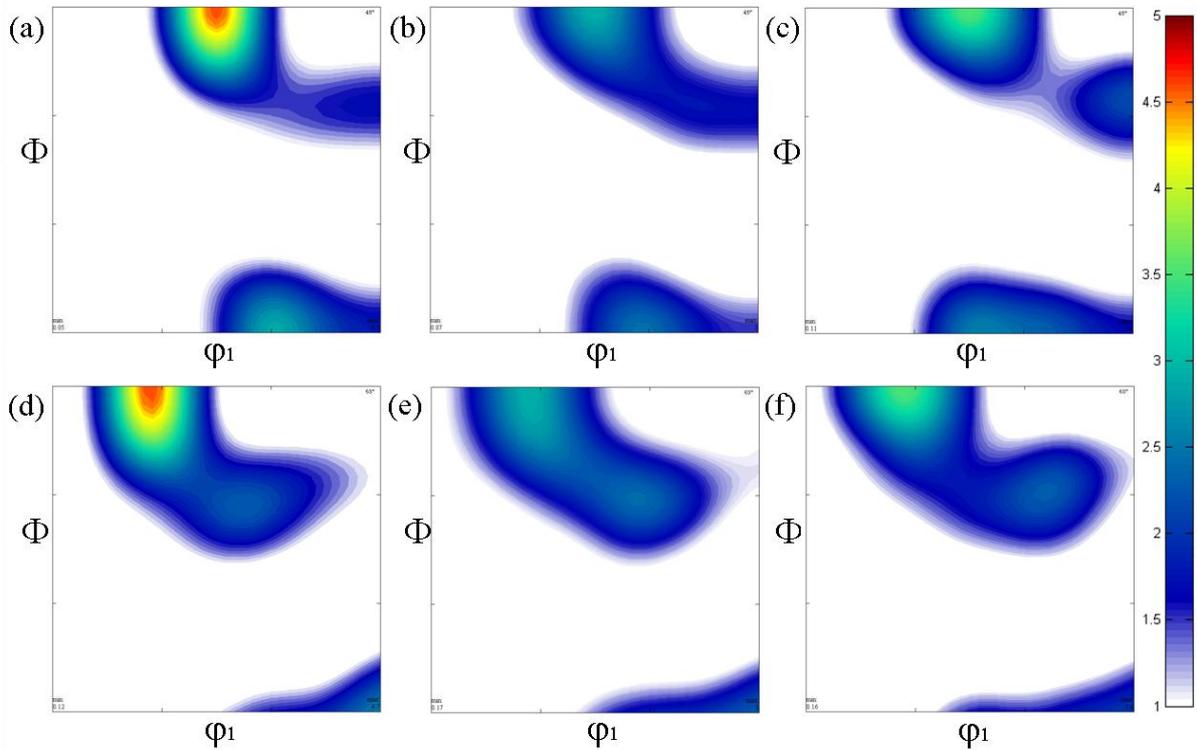

**Fig. 11.** ODF sections of $\gamma$ grains at (a-c) $\varphi_2 = 45°$ and (d-f) $\varphi_2 = 63°$ of samples 1, 2 and 3, respectively.

4.4. Pairing and selection of variants

Using both the $\alpha$ and $\gamma$ orientation mappings, the exact OR is calculated for each pixel and is subsequently assigned as a variant number (provided that the deviation from this variant is lower than 5°). The description of variants considered in the present study is given in Table 2. The K-S OR is used as the deviation angle between K-S, and the refined ORs are not large enough to affect the variant classification. The K-S variants are identified using the Bishop–Hill (B-H) [52] nomenclature, i.e., are named according to the CPs and CDs that correspond to the K–S parallelism conditions. However, the phenomenological theory of martensite crystallography [53] states that the CPs are not exactly parallel, with a misorientation of approximately 1°. As this deviation is on the order of the EBSD angular resolution, the use of the KS OR is reasonable and allows for an elegant classification scheme [54]. Note that the definitions of the CP and Bain [55] groups used in this work are given in the legend of Table 2.

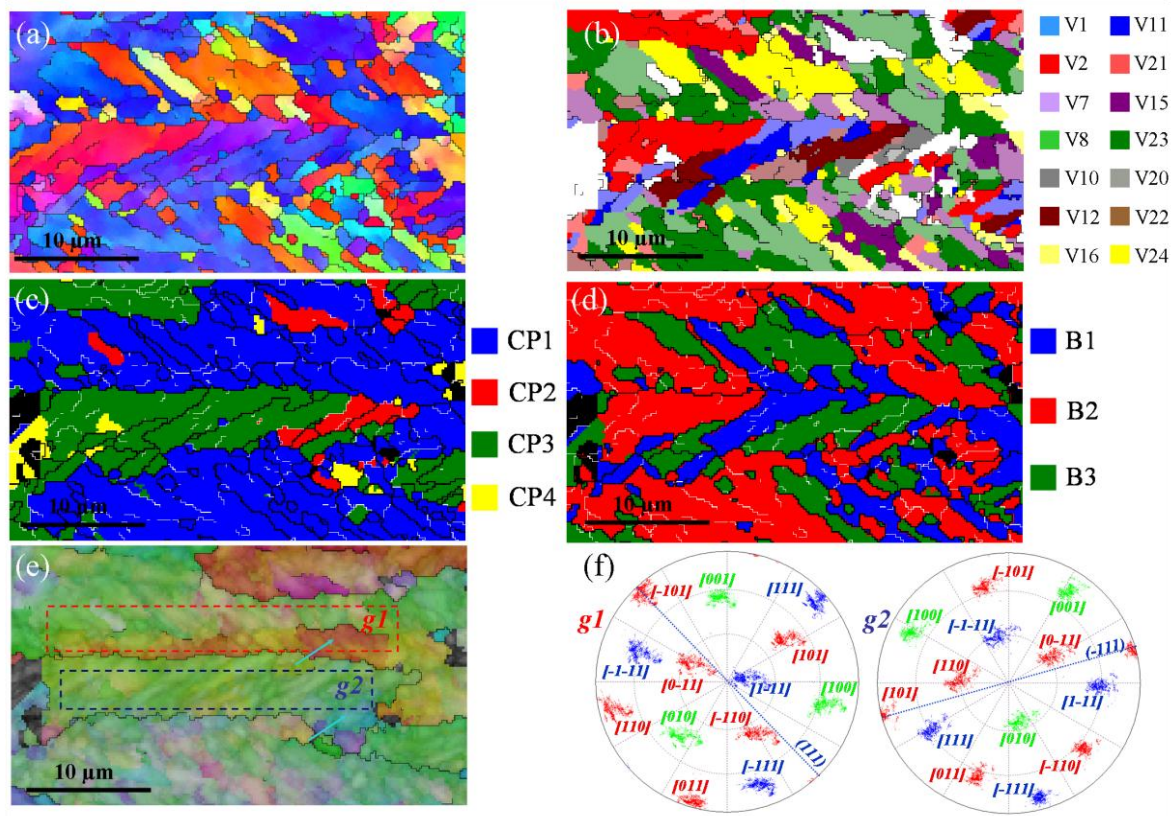

**Fig. 12.** (a) Experimental $\alpha$, (b) variant, (c) CP group, (d) Bain group and (e) reconstructed $\gamma$ orientation mappings of a selected subset of sample 1. Stereographic projections of $\gamma$ orientations in grains *g1* and *g2* are marked in (e).

A subset of the IPF mapping of sample 1 is selected and shown in Fig. 12(a); this subset is then coloured in Fig. 12(b) as a function of the variant number. Figs. 12(c-d) show the spatial distribution of the CP and Bain groups in this subset. The combination of the aforementioned mappings with the associated $\gamma$ reconstructed area, shown in Fig. 12(e), yields the following observation: one $\gamma$ grain generally consists of only one CP group, in which all three Bain domains are usually present. The same observation can be made for the other samples (see, e.g., Fig. 13 for sample 3). However, a difference can be observed in sample 1: the presence of a few twin-related variant pairs, as shown in Fig. 14. Interestingly, only two Bain groups are present within the inherited packet.

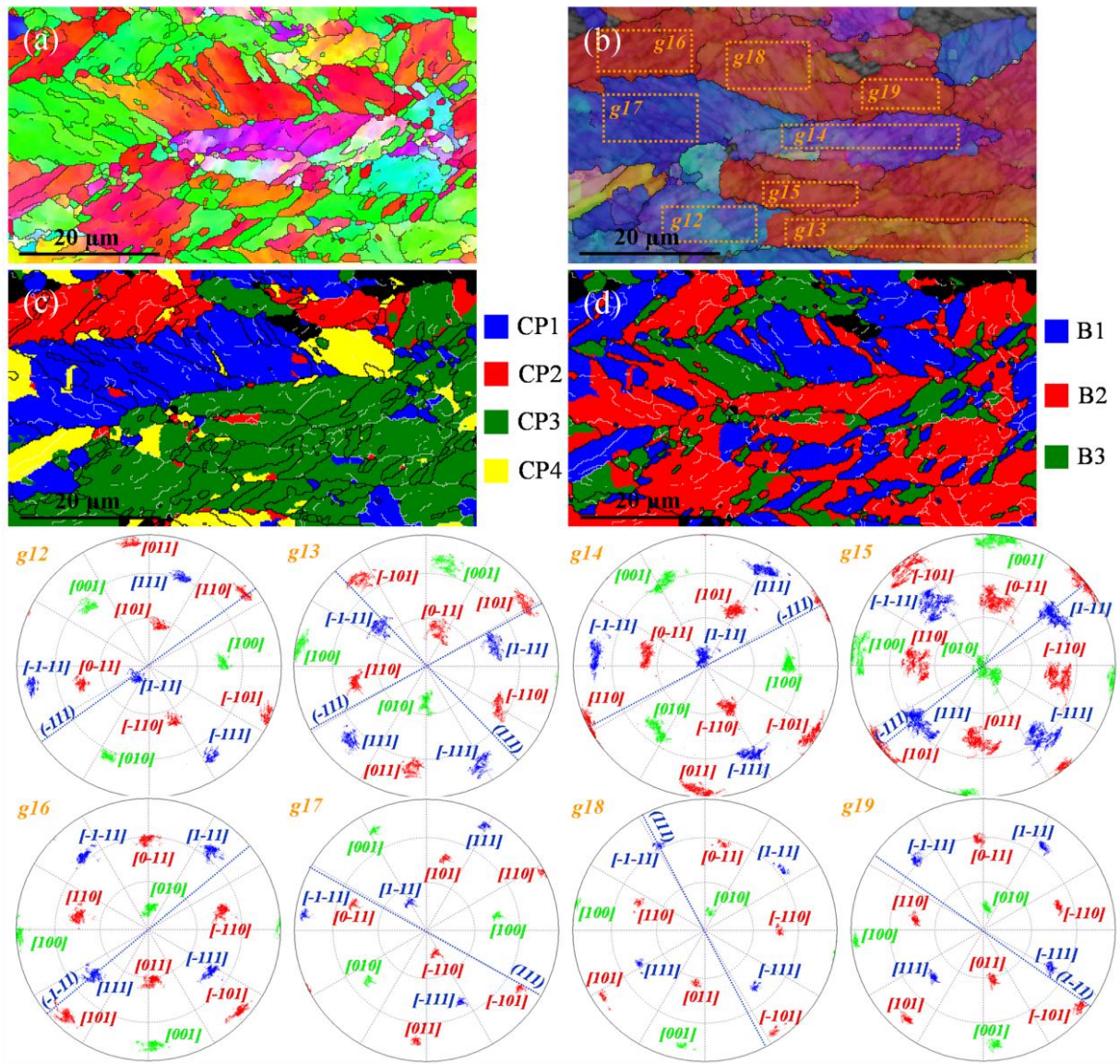

**Fig. 13.** (a) Experimental $\alpha$, (b) reconstructed $\gamma$, (c) CP group and (d) Bain group mappings of a selected subset of sample 1 showing twin-related pair variants in one packet. Stereographic projections of $\gamma$ orientations in grain *g3* are marked in (b).

In addition, the habit planes of lath martensite, i.e., $\{575\}_\gamma$, are known to be nearly parallel to the $\{111\}_\gamma$ planes shared by both the parent $\gamma$ grain and the inherited variant [56-58]. The stereographic projections shown in Figs. 12-13 indicate that the traces of $\{111\}_\gamma$ planes in the parallel relation between $\alpha$ and $\gamma$ (e.g., the $(-111)_\gamma$ plane from the CP3 group of grain *g12* in Fig. 13) are approximately parallel to the variant boundaries for all the $\gamma$ grains. Therefore, the habit plane, which is common to all laths in one CP group, is confirmed to systematically correspond to the $\{111\}_\gamma$ plane shared by both $\alpha$ and $\gamma$ grains.

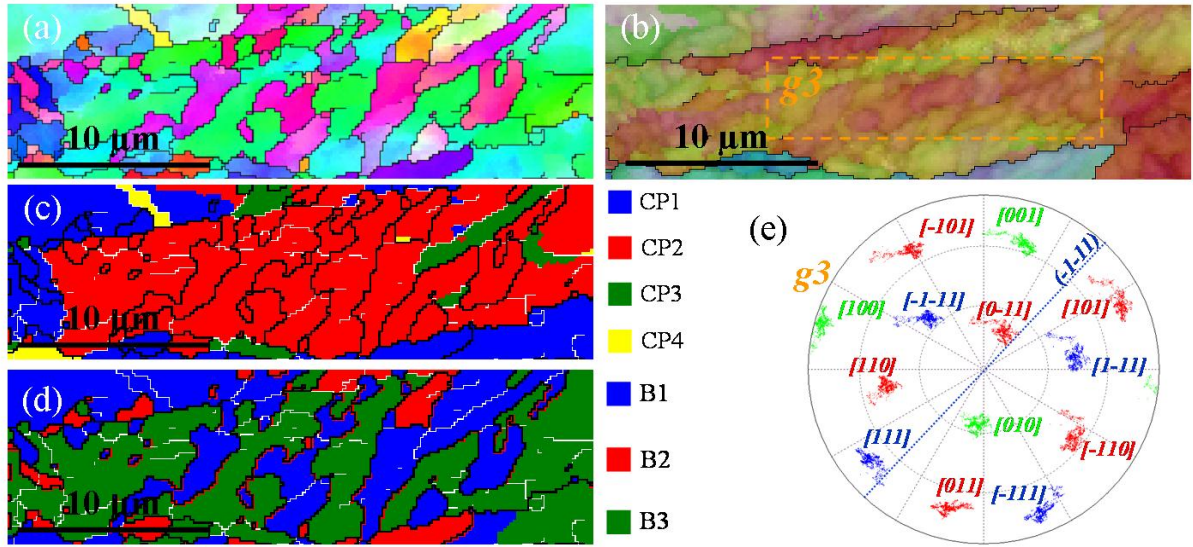

**Fig. 14.** (a) Experimental $\alpha$, (b) reconstructed $\gamma$, (c) CP group and (d) Bain group mappings of a selected subset of sample 3. Stereographic projections of $\gamma$ orientations in grains *g12* and *g19* are marked in (e).

Table 2 gives the misorientation angles of all the variants with respect to the first one (noted V1) using either the K-S OR or the refined ORs measured on the three samples. Two variants, V4 and V11, are equally close to V1 when considering the K-S OR, with a disorientation angle of ~10.5°. However, when the refined ORs are employed, V11 becomes much closer to V1, with a misorientation angle of 5-8°. This variant, which is the closest one to V1 regardless of the selected OR, belongs to both the same CP and Bain group as V1. The list of closest variants is given in the last column of Table 1. These variants only differ in brightness in Fig. 12(b), e.g., light blue for V1 and dark blue for V11; this colour coding allows for a better visualisation of their relative positions. As seen from Fig. 12, each Bain group systematically contains the pair of closest variants, which are, however, formed in varying amounts (note, e.g., the higher fraction of V2 compared to V21 in the Bain group B2 of grain *g2*).

A statistical quantification of the variant fractions can be derived from the high number of orientation data available in the IPF mappings of $\alpha$ and reconstructed $\gamma$ shown in Fig. 7. The reconstruction process fails to univocally determine the $\gamma$ orientation at some locations; these "unidentified" $\gamma$ areas are displayed in black in the IPF mappings of Figs. 7(d-f). A detailed comparison with SEM images after Nital etching indicates that the unidentified zones correspond to the M/A bands present in the analysed steels. The presence of these bands is clearly visible in Figs. 6(a-c). In summary, the $\gamma$ structure is not recovered from most of the M/A bands, implying that the statistical VS analysis shown in the following is almost exclusively related to the bainitic microstructure.

The statistical variant analysis for the complete orientation data set is shown in Fig. 15(a). The equipartition line at ~ 4.17% (dashed line) acts as a guideline for better visualisation. The results indicate that the group constituted by the first eight variants stands above the equipartition line in all three rolling conditions. As seen from Table 1, these eight variants share the same Burgers vector (i.e., ±III) according to the B-H system. This originates from the VS of $\alpha$ grains inherited from cube-oriented $\gamma$ grains, as seen from Fig. 15(b). It can be shown that the $\gamma$ grains with deformed texture components do not favour $\alpha$ grains with a particular type of Burgers vectors.

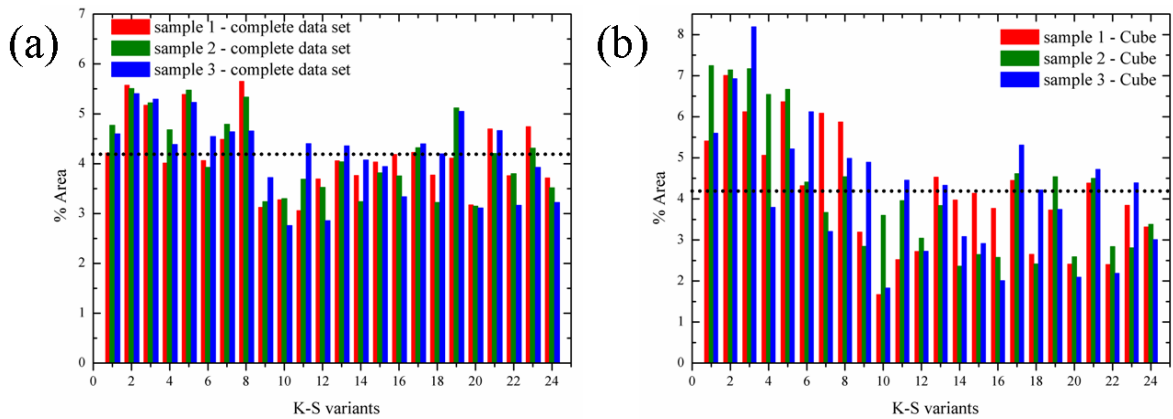

**Fig. 15.** Area fraction of variants as a function of their K-S identification number (given in Table 1) inherited from (a) all the $\gamma$ grains and (b) the parent cube-oriented grains.

Figs. 16(a-f) illustrate the VS resulting from various $\gamma$ orientations, with variants grouped into CP or Bain domains according to Table 2. We extract each texture component for the VS analysis by considering grains misoriented from the ideal component by less than 15°. Fig. 16(e) shows an almost perfect equipartition between the four CP domains, as all the area fractions range between 23 and 27% regardless of the CP group or the rolling condition. However, as seen from Figs. 16(a-d), the relative proportions between CP groups are more pronounced when considering each texture component separately, which may be explained by various slip activities on $\{111\}_\gamma$ planes as a function of the grain orientations. To further investigate this assumption, the predicted shears of all $\{111\}_\gamma$ planes for each parent

orientation (calculated using a full-constraints Taylor simulation [59]) are shown in Table 3 and compared to the VS plots in Figs. 16(a-d).

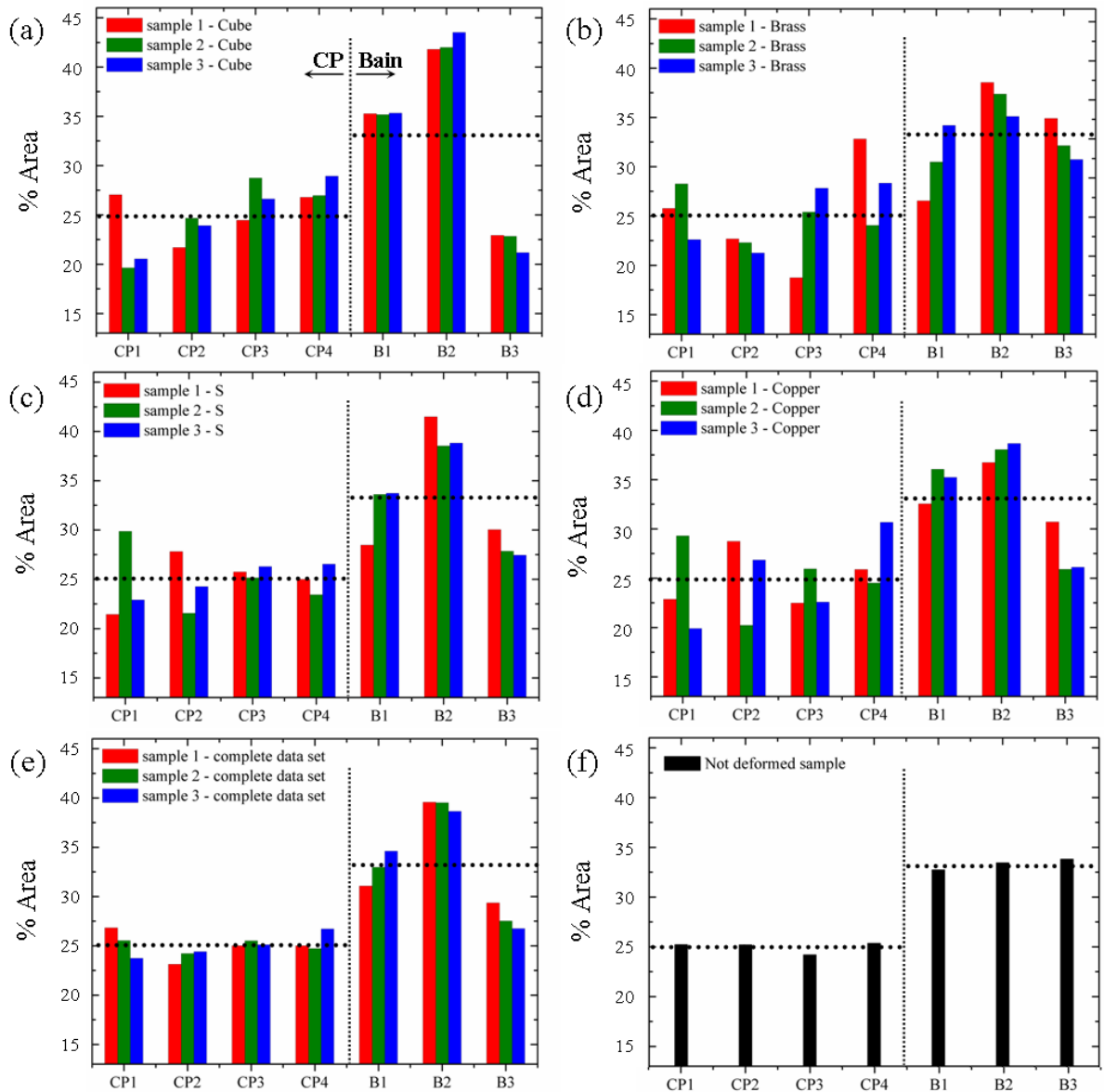

**Fig. 16.** Area fraction of CP and Bain variants inherited from (a) parent cube, (b) parent brass, (c) parent S, (d) parent copper and (e) all the parent grains in samples 1, 2 and 3. (e) Area fraction of CP and Bain variants inherited from parent grains in the non-deformed sample.

It should be noticed that Goss grains are not included in this statistical analysis as they cover less than 6% of the area in the $\gamma$ mappings. The results show that ~25% of the bainitic area inherited from parent copper or S grains is formed by variants belonging to the CP2 group whose predicted shear is null. The same observation can be made for the CP4 group from brass-oriented grains. Therefore, the observed proportions of CP groups are not accurately predicted by the calculated shears. In contrast, the proportions of Bain groups show some distinguishing features, e.g., the predominance of the B2 group for all the texture components, while the B3 group is observed less frequently. Note also that a lower FRT generally leads to more type-B1 and fewer type-B3 variants. Fig. 16(f) shows that there is no preferential Bain domain for the non-deformed sample.

| Orientation | Cube | Copper | S | Brass | Goss |
|---|---|---|---|---|---|
| CP1 (111) | 0.66 | 1.48 | 1.42 | 1.24 | 1.26 |
| CP2 (-1-11) | -0.59 | 0.00 | 0.00 | 0.80 | 0.00 |
| CP3 (-111) | -0.66 | -0.99 | -0.79 | -1.20 | -1.24 |
| CP4 (1-11) | 0.62 | -0.98 | -1.12 | 0.00 | 0.00 |

**Table 3.** Predicted shears of all $\{111\}_\gamma$ CPs for each parent orientation (calculated using a full-constraints Taylor model).

A straightforward way of visualising the effect of VS on the final texture is to plot the ODF sections of $\alpha$ grains inherited from each parent orientation. In addition, these measured $\alpha$ ODF sections can be easily compared to those simulated without any VS. This work is accomplished in Fig. 17. First, it can be noted that the VS of cube-oriented $\gamma$ grains is fairly similar for the three samples; they all exhibit a higher rotated Goss intensity. Similarly, the VS from copper $\gamma$ grains is rather insensitive to the FRT, as it systematically promotes the I(112)[1-10] component against the rotated Goss one. In contrast, the three other parent orientations generate substantially different bainite textures as a function of the rolling condition. Regarding the parent Goss, the intensities of the inherited E'(111)[0-11] and (223)(1-10) components vary significantly depending on the material; a low FRT promotes the latter orientation. Note that the fairly strong E' component inherited from $\gamma$ Goss grains in sample 3 accounts for the continuous aspect of the ND//<111> fibre observed in this sample, as seen from Fig. 10(c). Conversely, the weaker E' orientation in the other samples is responsible for truncating the aforementioned fibre. The XRD ODF sections, shown in Figs. 10(d-e), confirm that the intensity of the E' component is slightly higher in sample 3 than in sample 1. The VS applied to the parent brass and S grains follows roughly the same trends, i.e., a low FRT results in an intensity increase of the (332)[-1-13] orientation and, to a lesser extent, the F(111)[1-21] orientation. The rotated cube orientations inherited from parent brass and S grains are generally weak, except for those inherited from the parent brass in sample 3.

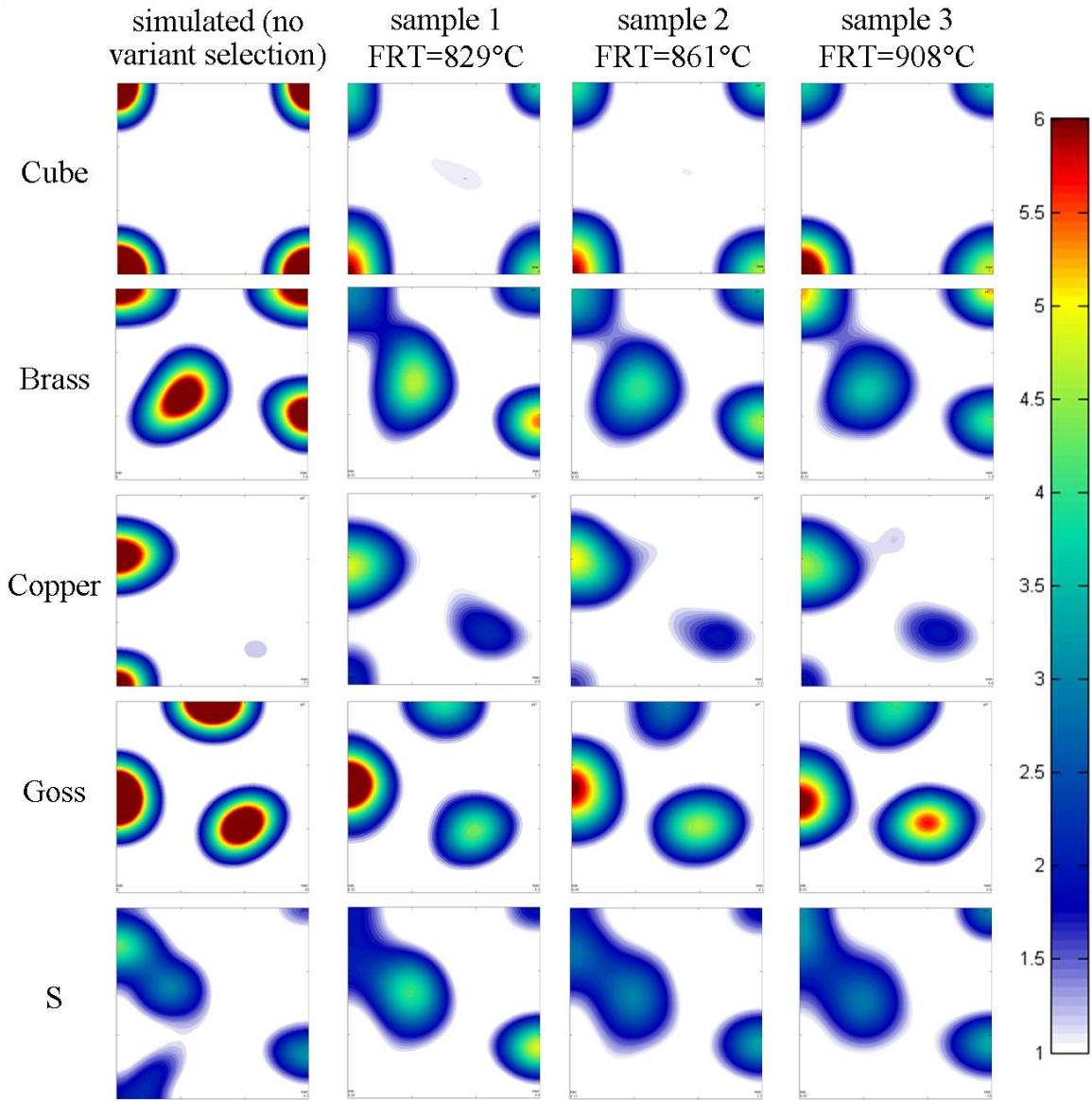

**Fig. 17.** Simulated without any variant selection (first column) and measured (2$^{nd}$ to 4$^{th}$ columns) ODF sections at $\varphi_2 = 45°$ of $\alpha$ grains inherited from different $\gamma$ orientations.

## 5. Discussion

5.1. Description of the bainite microtexture

The microstructure of the three specimens is essentially composed of lower bainite, as described above [48]. The variant pairing is found to be similar for the three samples: one prior $\gamma$ grain usually transforms into only one packet, which consists of a group of laths holding the same parallel CP relationship in the K-S OR. These CPs are found to be parallel to the habit planes. One packet is generally partitioned into three crystallographically distinct blocks, each of which is composed of two interleaved variants separated by a low misorientation angle, such as the pair V1/V11 shown in Table 1. These variants, or sub-blocks

[57], are the two K-S relations that share the same Bain axis of the transformation. Therefore, the present microstructures are similar to that of standard low-carbon lath martensite [57,60], but with the main difference being the restricted number of packets inherited from one prior $\gamma$ grain.

The preferential development of a single packet in a parent grain has already been observed by Malet et al. [61] in a bainite 0.4 C, 1.5 Si, 1.5 Mn (wt.%) transformation-induced plasticity (TRIP) steel deformed to an effective strain of $\varepsilon = 0.8$ at 1023 K. In contrast, the four CP groups were almost equally represented in the same TRIP steel deformed at $\varepsilon = 0.2$ [61]. Therefore, the large amount of deformation most likely accounts for the single-packet growth. Furthermore, the SEM image in Fig. 6(d) shows that most of the bainite grain boundaries are orientated at ~20-40° to the rolling direction (RD), in agreement with the $\gamma$ CP planes parallel to the habit planes (as shown in Figs. 12-13). The traces of the active {111} slip planes in $\gamma$, based on the Schmid factor analysis, are known to be oriented from 19.5° to 45° from the rolling direction for the main fcc texture components (i.e., brass, S, Goss, copper and cube) [62]. Because the dislocations accumulated along slip bands are supposed to promote the nucleation of $\alpha$ variants [16,63-64], it can be reasonably assumed that the high strain promotes the accumulation of dislocations on one particular {111} plane, whereas for lower strains, the dislocations are more equally distributed among the slip planes having the highest Schmid factors. The $\gamma$ grains that transform into more than one packet (e.g., the grain *g13* in Fig. 13) are most likely locally experiencing a lower plastic strain. Another relevant factor is the influence of the $\gamma$ grain size, which is relatively small in the present samples (< 15 μm). Then, the first nucleated variant may easily dominate in a single $\gamma$ grain, thus reducing the total number of variants inherited from one $\gamma$ grain.

Takayama et al. [65] reported that the microstructure of a bainitic low-C steel isothermally transformed at 450°C, referred to as "Type II" by the latter authors in contrast to that of lath martensite, referred to as "Type III", consists of ordinary packets in which the twin-related pairs are dominant, such as the V1/V2 pair shown in Table 2. Although several $\gamma$ grains transform into this Type II arrangement in sample 1, their contribution remains negligible compared to that of Type III. However, the latter authors [65] indicated that the Type II and Type III structures can be produced under approximately the same set of transformation temperatures and chemical driving forces. In addition, the nanobainite steel investigated in [65] contains 0.2 wt.% Si, which most likely implies a substantially different chemical driving force and a suppression of carbide growth compared to the analysed C-Mn bainitic steels.

Therefore, the present results confirm that the transition from Type II to Type III structures is very sharp and that the variant pairing of lath martensite can be generated in steels almost isothermally transformed above Ms during coil cooling. Furthermore, despite the sharp Type II/Type III transition, the presence of Type II $\gamma$ grains in sample 1 indicates that a gradual transition between the two latter structures is possible at intermediate temperatures.

The $\alpha$ texture of sample 1 is much sharper than that of samples 2 and 3, as observed in Fig. 10. Moreover, the predominant component in sample 1 is the {332}<-1-13> one. These results are in general agreement with a previous work [29] reporting an increase in the texture sharpness in plain C with and without Nb when decreasing the FRT. The latter authors also reported a transition from the {113}<110> and {332}<-1-13> components to the {223}<110> and {554}<225> ones as the FRT decreases due to rolling in the $\alpha+\gamma$ intercritical range. The absence of such a transition in the present samples confirms that the microstructure is fully austenitic shortly after hot rolling. This finding is of importance, as the crystallographic reconstruction process by definition cannot be applied to rolled $\alpha$ structures; there is indeed the assumption that the $\alpha$ orientation depends only on the $\gamma$ orientation and the $\gamma$-to-$\alpha$ phase transformation.

5.2. Effect of FRT on the $\gamma$ conditioning

As seen from Figs. 8(a-c), samples 1 and 2 consist of similar elongated "pancake" $\gamma$ grains, even though their FRTs are different. In the case of sample 3, the temperature was higher, and as a consequence, the structure was partially recrystallised. This result is consistent with the reported minimal effect of FRT at low temperatures (typically < 1000°C) on the $\gamma$ grain size in microalloyed steels [66]. This is a consequence of the similar amount of deformation both materials 1 and 2 received after the last recrystallisation in the hot rolling process. The recrystallisation rate can be evaluated from the GAM profiles presented in Fig. 8(e). The non-deformed steel shows a constant GAM of ~ 3.5°, which is the result of both the EBSD experimental error (e.g., noise, angular resolution, step size) and the plastic accommodation induced by transformation strain. This value of 3.5° is quite similar to the value of 3° reported elsewhere [25]. Fig. 8(e) indicates high GAM values (up to 10°) for the largest $\gamma$ grains in samples 1 and 2. The slopes after the linear regression fitting of GAM plots shown in Fig. 8(f) are equal to zero for the non-deformed sample, 2.2 for sample 3 and 3.4 for samples 1 and 2. These result confirm that (i) the contribution of rolling to the $\gamma$ misorientation is much more significant than that of the plasticity induced by the $\gamma$-to-$\alpha$ transformation [25] and (ii) the hot deformation is similar in samples 1 and 2, which results in similar $\gamma$ grain sizes. Note that the latter slopes might serve as an accurate parameter for assessing the recrystallisation fraction in hot-rolled steels. This possibility must be confirmed using mechanical methods, such as double-hit and stress-relaxation tests [67].

The smaller $\alpha$ grain size for samples 1 and 2 compared to sample 3 is a direct consequence of the difference in $\gamma$ grain size, as commonly observed in the literature [68]. As shown in Fig. 8, the block and packet (taken as the prior $\gamma$ grain, as we observed that the latter generally transforms into only one packet) sizes in samples 1 and 2, which are smaller than those in sample 3, are fairly comparable. The effective grain size for strength in lath martensite is reported to be both the packet size (as it relates to the mean free path of dislocation glide along {110} planes) [69] and the block size (whose effect depends on the cleanliness along the boundaries) [70]. As seen in Table 1, the tensile strength of samples 1 and 2 are similar, in agreement with the block and packet size considering the Hall-Petch effect. In contrast, the yield strength of sample 1 (1010 MPa) is larger than that of sample 2 (961 MPa). Therefore, the grain refinement is not the main cause of the improvement of yield strength at a low FRT. The beneficial effect of the {332}<-1-13> texture component on yield strength [29] may be invoked to account for the latter observation, but other microstructure parameters (e.g., precipitation, boundary decoration by carbon, dislocation density, secondary phases) must be considered for a quantitative correlation between the inherited microstructure and the mechanical properties, which is beyond the scope of this work.

Fig. 11 reveals that the $\gamma$ texture of the three samples is fairly similar, with the only difference being a slight increase in the cube component in sample 1. Of particular interest, the intensity of the deformation components (along both the brass-Goss and the brass-S-Cu fibres) is comparable in all the samples. This result indicates that the sharper inherited texture in sample 1 is mainly due to a different VS mechanism, which will be discussed in more detail in section 5.3.

Similar textures, observed for samples undergoing different recrystallisation rates, are somewhat unexpected in view of the well-known [18,50-51] recrystallised and "pancake" texture components in steels. However, a recent study [21] on stainless steel and Ni–30 wt.% Fe model alloys reported no significant change in the $\gamma$ texture character with increasing deformation temperature. The main reason for this lack of change lies in an analogous texture for the recrystallised grains and the deformed matrix. The latter authors [21] explain this observation by a GB migration of the deformed grains during recrystallisation and a deformation of recrystallised grains during their growth. Furthermore, the substantial amount of solute elements, such as manganese, and the short interpass time between rolling steps retard and limit the static recrystallisation process in the present case [50-51].

Taylor et al. [21] also reported that the cube texture component is an important part of the orientation distribution at high deformation temperatures, especially for stainless steel. In the present study, however, the strongest cube component is measured for sample 1, i.e., at the lowest FRT. The explanation relates to the strain accumulated prior to recrystallisation, which is known to strengthen the cube component [18]. Given that the initial thickness (35 mm), final thickness (4 mm), rolling speed and inter-stand distance are identical for the three samples, a lower deformation temperature at the final stand has been achieved by using lower deformation temperatures along the processing route. For the same thickness reduction, the applied force in the first stand was indeed measured to be larger for sample 1, leading to a higher accumulated strain. This effect produces a much sharper cube component during the recrystallisation process in the second stand. The comparison between the strip thickness at the exit of the first stand (~ 22 mm) and the final grain aspect ratio (~ 3.5, as observed in Fig. 8(c)) confirms that the sample has most likely been recrystallised in the second stand. Once formed, the cube texture remains very stable due to its low stored energy [71-72]. Nevertheless, it must be kept in mind that the temperature level with respect to $T_{nr}$ also affects the strength of the cube component. As the deformation temperatures are higher in sample 3 along the complete processing route, the recrystallisation in the latter sample is thermally activated in more stands, which may lead to an increased cube fraction as well. As a consequence, the prediction of the cube component intensity in multi-pass hot-rolled C-Mn steels is not straightforward, as it is not directly proportional to the temperature due to the competing accumulated strain and thermal activation effects.

5.3. Effect of FRT on the variant selection

In the previous section, it has been shown that the VS alone can account for the observed sharper inherited texture in the sample finish rolled at 829°C. It must be noted that this result is in contradiction with those of Shirzadi et al. [32] and Gong et al. [33], who recently reported that ausforming at 600°C, unlike ausforming at 300°C, does not activate any VS. However, many factors influence the VS mechanism [e.g., 73], and care must be taken when comparing the collected data. For instance, Gong et al. [33] investigated the VS in nanobainite steels (0.79C, 1.51Si; wt.%) ausformed in a hot compression machine on cylindrical specimens. The present study addresses low-carbon (0.15C wt.%) low-alloy steels deformed under plane strain compression (PSC). The latter authors [33] reported that the higher strength of the VS is due an increase of the transformation temperature, i.e., a lower chemical driving force [27]. Therefore, it is interesting to discuss the present results in terms of transformation temperature.

First, as discussed above in section 5.1, the presence of Type II grains in sample 1 indicates a transition microstructure between lath martensite and bainite formed at 450°C, unlike the pure Type III character in the other samples. This observation is consistent with an increase of transformation temperature [65]. Second, Stormvinter et al. [74] observed a shift from K-S to near G-T OR as the carbon content decreases (i.e., corresponding to a higher transformation temperature). The refined ORs, shown in Fig. 9, exhibit a higher deviation angle between the $\alpha$ and $\gamma$ CDs as the FRT is decreased. Again, this result implies a higher transformation temperature in sample 1. This reasoning assumes that the effects of other parameters affecting the OR, e.g., $\gamma$ strength for accommodation of the shape strain, are constant. Third, the GAM profiles of the $\alpha$ grains, plotted in Fig. 8(e), show a lower misorientation in sample 1 than in the other samples. Considering that the dislocations are associated with variations in orientation within each grain, the latter result suggests a lower dislocation density in sample 1. As the bainite transformation occurs in a temperature range where the shape change cannot be elastically accommodated, it is well-known that the required plastic deformation generates a build-up of dislocations, increasing in number as the transformation temperature decreases [75-76]. Therefore, a high transformation temperature entails a low dislocation density in the inherited microstructure, which is the case for sample 1. The last three aspects suggest that the strong VS taking place at a low FRT is due to an increase in the transformation temperature.

Fig. 16 reveals the growth of preferred Bain groups as a function of the FRT. These results highlight the importance of Bain groups in the VS mechanism. The Bain variants are imaginary and are never observed in reality, but they play a major role in the variant pairing, as previously noted. Their most significant feature is the Bain axis i.e., the $\gamma$ cube axis <100>, which is compressed to generate the $\alpha$ variant (a compression expressed as the Bain strain [77]). However, Fig. 18 indicates that the B2 and B3 Bain domains are the most and least dominant Bain groups, respectively, regardless of the parent grain orientation. As the <100> directions are differently oriented with respect to RD and ND (i.e., the two major deformation directions considering plane-strain compression) for each parent orientation, it follows that the Bain strain [77] has no or little influence on the VS. Kundu et al. [78] reported the same conclusion by showing that the close proximity of the variants in one Bain group and their nonparallel orientation play a more predominant role in VS than the Bain strain itself.

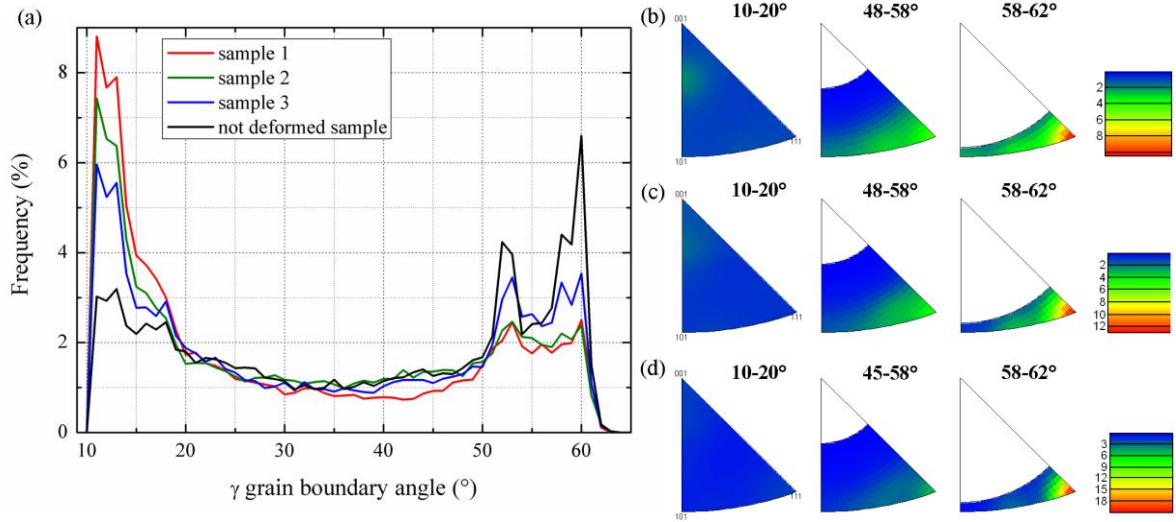

**Fig. 18.** Grain boundary misorientation: (a) angle distribution profiles, (b-d) axes for samples 1, 2 and non-deformed, respectively.

The equal contributions of the three Bain groups in the non-deformed steel, as seen in Fig. 16(f), offers an alternative explanation for the occurrence of preferential Bain variants in the hot-rolled sample. It is well-known [70] that the packet preferentially contains equal fractions of the three Bain blocks, as this arrangement minimises the elastic energy (the net transformation strain being almost a simple dilatation). In the case of a hot deformation, the polygranular matrix must accommodate the deformation by a modification of several interrelated microstructure parameters, especially the grain shape and the type of grain boundaries (GBs). Fig. 18(a) shows that the proportion of GB misorientation angles higher than 45° decreases as the FRT is increased. In addition, as seen from Figs. 17(b-d), the misorientation axes of these GBs are only of type <111> in the non-deformed sample, whereas there is a spread of GB axes from <111> to nearly <110> in the rolled samples.

It has been recently shown [79-82] that the strain accommodation may take place across the parent grain boundary by selecting some preferred variants that hold a near K-S OR with the two adjacent $\gamma$ grains. Let us apply this VS criterion to the observed modification of GB axes in the non-deformed and hot-rolled samples. We denote by *g1* and *g2* two arbitrary parent grains, which are separated by a GB of type 55°/<111>, as commonly observed in the non-deformed sample. It can be demonstrated that the following reasoning can be applied to all the GBs with a misorientation angle larger than ~ 45°. We denote by V1 to V24 the 24 variants inherited from *g1* using the K-S OR. The disorientation angles from the K-S OR with the parent grain *g2* that hold the variants Vi are given in Table 4.

| GB between γ grains g1 and g2 | Disorientation angle from the K-S OR with γ grain g2 that holds a variant Vi inherited from γ grain g1 (degree) | | | | | | | | | | | | | | | | | | | | | | | | Mean disorientations/Bain group | | |
|---|---|---|---|---|---|---|---|---|---|---|---|---|---|---|---|---|---|---|---|---|---|---|---|---|---|---|---|
| | V1 | V2 | V3 | V4 | V5 | V6 | V7 | V8 | V9 | V10 | V11 | V12 | V13 | V14 | V15 | V16 | V17 | V18 | V19 | V20 | V21 | V22 | V23 | V24 | B1 | B2 | B3 |
| [-111]/55° | 20.18 | 17.04 | 20.15 | 14.22 | 11.65 | 11.65 | 5.00 | 5.00 | 17.04 | 20.18 | 20.15 | 14.22 | 11.65 | 11.65 | 5.00 | 5.00 | 20.18 | 17.04 | 14.22 | 20.15 | 11.65 | 11.65 | 5.00 | 5.00 | 13.11 | 13.11 | 13.11 |
| [-1-11]/55° | 5.00 | 5.00 | 11.65 | 11.65 | 20.18 | 17.04 | 20.15 | 14.22 | 11.65 | 11.65 | 5.00 | 5.00 | 14.22 | 20.15 | 20.18 | 17.04 | 17.04 | 20.18 | 20.15 | 14.22 | 5.00 | 5.00 | 11.65 | 11.65 | 13.11 | 13.11 | 13.11 |
| [-1-11]/55° | 14.22 | 20.15 | 17.04 | 20.18 | 5.00 | 5.00 | 11.65 | 11.65 | 5.00 | 5.00 | 11.65 | 11.65 | 20.15 | 14.22 | 17.04 | 20.18 | 11.65 | 11.65 | 5.00 | 5.00 | 20.18 | 17.04 | 14.22 | 20.15 | 13.11 | 13.11 | 13.11 |
| [1-11]/55° | 11.65 | 11.65 | 5.00 | 5.00 | 14.22 | 20.15 | 17.04 | 20.18 | 20.18 | 17.04 | 14.22 | 20.15 | 5.00 | 5.00 | 11.65 | 11.65 | 5.00 | 5.00 | 11.65 | 11.65 | 17.04 | 20.18 | 20.15 | 14.22 | 13.11 | 13.11 | 13.11 |
| [11-1]/55° | 17.04 | 20.18 | 14.22 | 20.15 | 5.00 | 5.00 | 11.65 | 11.65 | 5.00 | 5.00 | 11.65 | 11.65 | 20.18 | 17.04 | 20.15 | 14.22 | 11.65 | 11.65 | 5.00 | 5.00 | 20.15 | 14.22 | 17.04 | 20.18 | 13.11 | 13.11 | 13.11 |
| [1-1-1]/55° | 5.00 | 5.00 | 11.65 | 11.65 | 20.15 | 14.22 | 20.18 | 17.04 | 11.65 | 11.65 | 5.00 | 5.00 | 17.04 | 20.18 | 20.15 | 14.22 | 14.22 | 20.15 | 20.18 | 17.04 | 5.00 | 5.00 | 11.65 | 11.65 | 13.11 | 13.11 | 13.11 |
| [-1-1-1]/55° | 20.15 | 14.22 | 20.18 | 17.04 | 11.65 | 11.65 | 5.00 | 5.00 | 14.22 | 20.15 | 20.18 | 17.04 | 11.65 | 11.65 | 5.00 | 5.00 | 20.15 | 14.22 | 17.04 | 20.18 | 11.65 | 11.65 | 5.00 | 5.00 | 13.11 | 13.11 | 13.11 |
| [-11-1]/55° | 11.65 | 11.65 | 5.00 | 5.00 | 17.04 | 20.18 | 14.22 | 20.15 | 20.15 | 14.22 | 17.04 | 20.18 | 5.00 | 5.00 | 11.65 | 11.65 | 5.00 | 5.00 | 11.65 | 11.65 | 14.22 | 20.15 | 20.18 | 17.04 | 13.11 | 13.11 | 13.11 |
| Average | | | | | | | | | | | | | | | | | | | | | | | | | 13.11 | 13.11 | 13.11 |

**Table 4.** Disorientation angle from the K-S OR with $\gamma$ grain *g2* containing a variant Vi inherited from $\gamma$ grain *g1*. The grains *g1* and *g2* are misoriented by an angle of 55° about the direction <111>.

It can be noticed that the mean disorientation angles are similar for all the Bain groups. Furthermore, the six most likely nucleated variants (i.e., those which deviated by only 5° from the K-S OR with *g2*) are equally present in the three Bain groups. The same calculation in the case of a 55°/<110> GB, as shown in Table 5, reveals a strong dependence of the mean deviation angle on the Bain group. For instance, GBs of type 55°/[hk0], with h=±1, k=±1, promote the growth of the B1 and B2 Bain variants, as the latter are deviated ~ 13° from the K-S OR of *g2* in contrast to a ~ 24.6° deviation for the B3 group. Therefore, the crystallography of the <110> GBs may play a significant role in promoting particular Bain groups. It can be shown that the same conclusion can be drawn for all axes comprised between the <110> and the <111> ones, such as <323> or <535>. In contrast, the growth of Bain domains is independent of the symmetry of GBs with a <111> misorientation axis.

| GB between γ grains g1 and g2 | Disorientation angle from the K-S OR with γ grain g2 that holds a variant Vi inherited from γ grain g1 (degree) | | | | | | | | | | | | | | | | | | | | | | | | Mean disorientations/Bain group | | |
|---|---|---|---|---|---|---|---|---|---|---|---|---|---|---|---|---|---|---|---|---|---|---|---|---|---|---|---|
| | V1 | V2 | V3 | V4 | V5 | V6 | V7 | V8 | V9 | V10 | V11 | V12 | V13 | V14 | V15 | V16 | V17 | V18 | V19 | V20 | V21 | V22 | V23 | V24 | B1 | B2 | B3 |
| [110]/55° | 5.00 | 5.00 | 5.00 | 5.00 | 20.18 | 19.15 | 20.18 | 19.15 | 17.04 | 22.69 | 11.65 | 15.53 | 11.65 | 30.44 | 14.22 | 29.71 | 11.65 | 30.44 | 14.22 | 29.71 | 11.65 | 15.53 | 17.04 | 22.69 | 12.99 | 12.99 | 24.59 |
| [-110]/55° | 19.15 | 20.18 | 19.15 | 20.18 | 5.00 | 5.00 | 5.00 | 5.00 | 11.65 | 15.53 | 17.04 | 22.69 | 14.22 | 29.71 | 11.65 | 30.44 | 17.04 | 22.69 | 11.65 | 15.53 | 14.22 | 29.71 | 11.65 | 30.44 | 12.99 | 12.99 | 24.59 |
| [-1-10]/55° | 5.00 | 5.00 | 5.00 | 5.00 | 19.15 | 20.18 | 19.15 | 20.18 | 14.22 | 29.71 | 11.65 | 30.44 | 11.65 | 15.53 | 17.04 | 22.69 | 11.65 | 15.53 | 17.04 | 22.69 | 11.65 | 30.44 | 14.22 | 29.71 | 12.99 | 12.99 | 24.59 |
| [1-10]/55° | 20.18 | 19.15 | 20.18 | 19.15 | 5.00 | 5.00 | 5.00 | 5.00 | 11.65 | 30.44 | 14.22 | 29.71 | 17.04 | 22.69 | 11.65 | 15.53 | 14.22 | 29.71 | 11.65 | 30.44 | 17.04 | 22.69 | 11.65 | 15.53 | 12.99 | 12.99 | 24.59 |
| [011]/55° | 22.69 | 17.04 | 11.65 | 15.53 | 11.65 | 30.44 | 29.71 | 14.22 | 30.44 | 11.65 | 29.71 | 14.22 | 15.53 | 11.65 | 22.69 | 17.04 | 5.00 | 5.00 | 5.00 | 5.00 | 19.15 | 20.18 | 20.18 | 19.15 | 24.59 | 12.99 | 12.99 |
| [0-11]/55° | 15.53 | 11.65 | 17.04 | 22.69 | 14.22 | 29.71 | 30.44 | 11.65 | 22.69 | 17.04 | 15.53 | 11.65 | 29.71 | 14.22 | 30.44 | 11.65 | 19.15 | 20.18 | 20.18 | 19.15 | 5.00 | 5.00 | 5.00 | 5.00 | 24.59 | 12.99 | 12.99 |
| [0-1-1]/55° | 29.71 | 14.22 | 11.65 | 30.44 | 17.04 | 22.69 | 22.69 | 11.65 | 15.53 | 11.65 | 22.69 | 17.04 | 30.44 | 11.65 | 29.71 | 14.22 | 5.00 | 5.00 | 5.00 | 5.00 | 20.18 | 19.15 | 19.15 | 20.18 | 24.59 | 12.99 | 12.99 |
| [01-1]/55° | 30.44 | 11.65 | 14.22 | 29.71 | 17.04 | 22.69 | 15.53 | 11.65 | 29.71 | 14.22 | 30.44 | 11.65 | 22.69 | 17.04 | 15.53 | 11.65 | 20.18 | 19.15 | 19.15 | 20.18 | 5.00 | 5.00 | 5.00 | 5.00 | 24.59 | 12.99 | 12.99 |
| [101]/55° | 11.65 | 30.44 | 29.71 | 14.22 | 15.53 | 17.04 | 22.69 | 11.65 | 5.00 | 5.00 | 5.00 | 5.00 | 20.18 | 19.15 | 19.15 | 20.18 | 11.65 | 30.44 | 11.65 | 29.71 | 14.22 | |
| [10-1]/55° | 17.04 | 22.69 | 15.53 | 11.65 | 29.71 | 14.22 | 11.65 | 30.44 | 20.18 | 19.15 | 19.15 | 20.18 | 5.00 | 5.00 | 5.00 | 5.00 | 15.53 | 11.65 | 22.69 | 17.04 | 29.71 | 14.22 | 30.44 | 11.65 | 12.99 | 24.59 | 12.99 |
| [-10-1]/55° | 11.65 | 15.53 | 22.69 | 17.04 | 30.44 | 11.65 | 14.22 | 29.71 | 5.00 | 5.00 | 5.00 | 5.00 | 19.15 | 20.18 | 20.18 | 19.15 | 29.71 | 14.22 | 30.44 | 11.65 | 15.53 | 11.65 | 22.69 | 17.04 | 12.99 | 24.59 | 12.99 |
| [-101]/55° | 14.22 | 29.71 | 30.44 | 11.65 | 22.69 | 17.04 | 11.65 | 15.53 | 19.15 | 20.18 | 20.18 | 19.15 | 5.00 | 5.00 | 5.00 | 5.00 | 30.44 | 11.65 | 29.71 | 14.22 | 22.69 | 17.04 | 15.53 | 11.65 | 12.99 | 24.59 | 12.99 |
| Average | | | | | | | | | | | | | | | | | | | | | | | | | 16.86 | 16.86 | 16.86 |

**Table 5.** Disorientation angle from the K-S OR with $\gamma$ grain *g2* containing a variant Vi inherited from $\gamma$ grain *g1*. The grains *g1* and *g2* are misoriented by an angle of 55° about the direction <110>.

Unlike the Bain domains, the four CP groups are fairly equally distributed for all the samples regardless of the parent orientation, as seen in Fig. 16. As the active slip planes are different for every $\gamma$ orientation (as shown in Table 3), this result implies that the habit planes of the selected variants do not systematically correspond to the predicted active $\gamma$ slip planes. A few parent grains are also examined in detail to confirm the latter finding. It can be noticed in

Table 6 that the CP planes of the highest Taylor factors are not observed in 13 out of the 19 grains investigated. Note also that 8 out of the 19 grains do not show one of the two most predictable CP planes. This result is in good agreement with a previous study by He [83] on the variant selection of hot-rolled TRIP steels. Using a full-constraints Taylor model, the latter author found a 70% agreement between the predicted and observed variants in only 15 out of 25 grains examined. These results reveal a noticeable lack of agreement between the observed and predicted CP planes on hot-rolled samples. Shear deformation during the hot rolling can not be invoked, as the $\gamma$ ODF sections of the analysed samples do not show any shear-type component (i.e., {001}<110> or {111}-fibre) [84], as observed in Fig. 7. Interestingly, Miyamoto et al. [85] and Gong et al. [33] recently reported the opposite conclusion on steels subjected to uniaxial compression. A plausible explanation may lie in the difference of deformation modes, which are known to generate substantially different textures [86] and grain shapes ($\varepsilon_{11}$=1 and $\varepsilon_{22}$=0 in plane-strain compression vs $\varepsilon_{11}$=$\varepsilon_{22}$=0.5 in uniaxial compression). These differences generate various polygranular accommodation mechanisms through, e.g., a change in boundary types, as observed in Fig. 18. In addition, the rates and kinetics of the recovery and recrytallisation processes may also vary as a function of the deformation modes (e.g., [87]). As discussed above, all these microstructure parameters play a significant role in the VS during the phase transformation. These results prove that the grain environment, at least in the steels deformed under plane-strain compression, must be taken into account for a reliable prediction of active $\gamma$ slip planes.

| Grains ID | Euler angles $\varphi_1/\Phi/\varphi_2$ (°) | CP1 (111) | CP2 (-1-11) | CP3 (-111) | CP4 (1-11) | Measured CP |
|---|---|---|---|---|---|---|
| g1 | 348.5/35.3/336.8 | 1.32 | 0.00 | 1.13 | 0.62 | CP1 |
| g2 | 56.6/37.2/314.6 | 1.48 | 0.46 | 1.33 | 0.00 | CP3 |
| g3 | 32.3/30.7/327.7 | 1.37 | 0.18 | 1.12 | 0.09 | CP2 |
| g4 | 184.9/38.8/176.1 | 0.00 | 1.29 | 0.00 | 1.25 | CP1 & CP2 |
| g5 | 238.9/34.2/149.5 | 0.71 | 1.16 | 0.00 | 1.42 | CP1 |
| g6 | 176.3/20.5/210.3 | 0.06 | 0.80 | 0.00 | 1.74 | CP2 |
| g7 | 217.8/27.1/167.4 | 0.50 | 1.17 | 0.00 | 1.43 | CP1 |
| g8 | 191.2/19.8/196.6 | 0.30 | 1.12 | 0.00 | 1.55 | CP4 |
| g9 | 54.4/41.8/268.8 | 0.00 | 1.21 | 1.14 | 1.16 | CP2 |
| g10 | 317.8/27.9/358.1 | 1.39 | 0.00 | 1.11 | 0.84 | CP4 |
| g11 | 250.8/24.6/143.7 | 0.82 | 1.10 | 0.00 | 1.48 | CP1 |
| g12 | 139.1/34.7/177.8 | 0.00 | 1.27 | 0.87 | 1.16 | CP3 |
| g13 | 18.8/20.8/342.2 | 1.26 | 0.08 | 1.05 | 0.13 | CP1 & CP3 |
| g14 | 320.0/42.7/3.8 | 1.24 | 0.00 | 1.20 | 0.77 | CP3 |
| g15 | 105.5/7.7/252.1 | 0.65 | 0.85 | 0.60 | 0.78 | CP3 |
| g16 | 188.4/14.8/175.6 | 0.14 | 1.09 | 0.21 | 1.07 | CP2 |
| g17 | 300.3/30.8/16.9 | 1.41 | 0.00 | 1.10 | 0.89 | CP1 |
| g18 | 134.4/16.0/228.1 | 0.43 | 0.87 | 0.57 | 1.13 | CP1 |
| g19 | 205.6/14.7/157.3 | 0.21 | 1.07 | 0.27 | 1.05 | CP4 |

**Table 6.** Comparison between the predicted shears of {111}$\gamma$ CPs and measured {111}$\gamma$ CPs in 19 grains randomly selected from samples 1, 2 and 3.

5.4. Orientation dependence of variant selection

The cube-oriented grains exhibit a particular VS (Fig. 15) based on the predominance of Burgers vectors of type ±III according to the B-H system. Taylor et al. [62] revealed a major difference in terms of substructures between the cube and non-cube orientations of the Ni-30Fe alloy, with the former being composed of subgrains filled with dislocation cells with low misorientations and the latter found to consist of elongated microbands containing fine subgrains. Therefore, the specific arrangement of the dislocations substructures within cube grains most likely accounts for their particular VS.

Fig. 17 shows that the strong intensity of ND//<111> and RD//<110> fibres in sample 1 results from a strong VS from parent brass/S- and Goss-oriented grains, respectively. In contrast, the VS from copper-oriented grains remains unchanged regardless of the FRT. The cause of the insensitivity of the VS from $\gamma$ copper grains to the rolling conditions remains unclear. Nevertheless, Taylor et al. [62], using a Ni–30Fe model alloy deformed in hot PSC, proved that the copper grains show the highest stored energy.

The driving force for GB migration during recrystallisation is proportional to the dislocation density difference across the GBs. Although the deformation temperature at the last stand is well below the $T_{nr}$ (as indicated by the pancaked microstructure), such recrystallisation processes may occur due to the large difference in dislocation density between the copper and cube grains. Indeed, the latter orientation undergoes fast dynamic and static recovery during and shortly after the rolling passes, which entails its low stored energy. Ridha et al. [87] explained this feature by noticing that only two types of Burgers vectors, which are orthogonal to one another, are active in cube grains during PSC.

In summary, the relatively low dependence of the VS from cube and copper grains with the FRT, as observed in Fig. 15, may be due to the start of the GB migration between the two latter grain orientations, even at low rolling temperatures. It has been reported [21] that the recrystallisation indeed occurs in Ni-30Fe alloys deformed at temperatures as low as 700°C. More investigation is needed to confirm this hypothesis.

6. Conclusion

First, an alternative approach for the crystallographic $\gamma$ reconstruction from EBSD mappings has been presented. It combines the strong points of two existing models: the OR refinement method, local pixel-by-pixel analysis and the nuclei identification and spreading strategy. This combination allows (*i*) the $\gamma$ orientation to be assigned with greater accuracy due to the use of a more severe criterion for the deviation angle between measured and predicted $\alpha$ orientations and an additional criterion ensuring the uniqueness of the $\gamma$ orientation, (*ii*) the $\alpha$ orientation gradient to be taken into account in the reconstruction process and (*iii*) the $\gamma$ grain orientation to be locally inferred. The method is then tested on measured data to prove the reliability of the crystallographic reconstruction approach. It is shown that the refined OR in a Q&P steel is representative of the true OR measured on retained $\gamma$ particles and that the reconstructed $\gamma$ microstructure agrees well with the one revealed by chemical etching on a lath martensite steel.

Second, the $\gamma$ reconstruction method is used for a detailed, statistical study of the effect of the FRT on the phase transformation characteristics in industrial C-Mn high-strength steels. Three samples, noted 1, 2 and 3, have been finish rolled at 829°C, 861°C and 908°C, respectively. The main results obtained in this study are as follows:

1. The three samples consist of lower bainite, with a similar variant pairing as standard low-carbon lath martensite (i.e., a prior $\gamma$ grain – packet – block – sub-block arrangement). The sample deformed at the lowest FRT also exhibits a few twin-related variant pairs, which reveals that both martensite and bainite structures co-exist at transformation temperatures just above Ms. In the present samples, one prior $\gamma$ grain is generally transformed into only one packet. It is also shown that the habit plane, which is common to all laths in one packet, corresponds systematically to the $\{111\}_\gamma$ plane shared by both $\alpha$ and $\gamma$ grains.

2. The GAM and grain sizes are similar in samples 1 and 2, despite the use of a different FRT. The misorientation induced by the hot deformation is confirmed to be much higher than that induced by the transformation strain. The use of the lowest FRT does not improve the tensile strength, which is similar for samples 1 and 2. This can be attributed to bainite microstructure refinement. In contrast, although the features of the bainitic lath structure are similar between both pancaked materials, the yield strength is higher for the lowest FRT. Among other factors, this could be due to the presence of the $\{332\}<-1-13>$ texture component.

3. The three samples show similar $\gamma$ orientation distributions. The only difference is a slight increase of the cube component in sample 1, which is due to a higher accumulated strain in the first stand, thus strengthening the cube orientation during recrystallisation in the second stand. The cube component intensity in multi-pass hot-rolled C-Mn steels is therefore not directly proportional to the rolling temperature. The intensities of the fcc deformation components are comparable in all the samples, which indicates that the sharper inherited texture observed only in sample 1 originates from a different VS mechanism.

4. The stronger VS measured in the sample finish rolled at the lowest temperature is most likely due to an increase in the transformation temperature, as supported by three microstructure aspects: the increase in twin-related pair variants, the lower dislocation density and an OR with a larger deviation angle between the $\alpha$ and $\gamma$ CDs in sample 1.

5. The strong intensity of ND//<111> and RD//<110> fibres in the transformation texture of sample 1 results from a strong VS from parent brass/S- and Goss-oriented grains, respectively. In contrast, the VS from copper-oriented grains remains unchanged regardless of the FRT. The difference in dislocation density between copper and cube grains is invoked to account for this behaviour.

6. It is shown that the grain environment, at least in the steels deformed under plane-strain compression, must be taken into account for a reliable prediction of active $\gamma$ slip planes using the Taylor model. In addition, a clear dependence of the Bain group frequency on the FRT is revealed regardless of the parent orientation. The Bain strain itself has no or little influence on this correlation. In contrast, an interpretation is given in view of the development of high-angle GBs with misorientation axes between the <111> and <110> directions in the rolled samples.